\documentclass[journal]{IEEEtran}

\usepackage{cite,graphicx,amsmath,amssymb}
\usepackage{subfigure}
\usepackage{citesort}
\usepackage{fancyhdr}
\usepackage{mdwmath}
\usepackage{mdwtab}
\usepackage{balance}
\usepackage{xcolor}
\usepackage{bm}
\usepackage{amssymb}
\usepackage{amsmath}
\usepackage{cite}
\usepackage{url}
\usepackage{xcolor}
\usepackage{cite,graphicx,amsmath,amssymb}
\usepackage{subfigure}
\usepackage{citesort}
\usepackage{fancyhdr}
\usepackage{mdwmath}
\usepackage{mdwtab}
\usepackage{caption}
\usepackage{amsthm}
\usepackage{algorithm}
\usepackage{algorithmic}
\usepackage{multirow}
\usepackage{flafter}
\usepackage{graphicx}
\usepackage{subfigure}
\usepackage{algorithm}
\usepackage{algorithmic}
\newtheorem{proposition}{Proposition}
\newtheorem{remark}{Remark}

\usepackage{amsmath,amsthm,amssymb}
\newtheorem{definition}{Definition}

\newtheorem{theorem}{Theorem}
\newtheorem{lemma}{Lemma}

\makeatletter
\def\ScaleIfNeeded{%
\ifdim\Gin@nat@width>\linewidth \linewidth \else \Gin@nat@width
\fi } \makeatother

\begin{document}

\title{Reinforcement Learning in Multiple-UAV Networks: Deployment and Movement Design}

\author{
Xiao~Liu,~\IEEEmembership{Student Member,~IEEE,}
 Yuanwei~Liu,~\IEEEmembership{Senior Member,~IEEE,}\\
        and Yue~Chen,~\IEEEmembership{Senior Member,~IEEE,}

\thanks{

X. Liu, Y. Liu, and Y. Chen are with Queen Mary University of London, London E1 4NS, UK. (email: x.liu@qmul.ac.uk; yuanwei.liu@qmul.ac.uk; yue.chen@qmul.ac.uk).

Part of this work was presented in the IEEE Global Communication Conference Workshops 2018~\cite{liu2018deployment}.
}
}

 \maketitle

\vspace{-2cm}

\begin{abstract}
A novel framework is proposed for quality of experience (QoE)-driven deployment and dynamic movement of multiple unmanned aerial vehicles (UAVs). The problem of joint non-convex three-dimensional (3D) deployment and dynamic movement of the UAVs is formulated
for maximizing the sum mean opinion score (MOS) of ground users, which is proved to be NP-hard. In the aim of solving this pertinent problem, a three-step approach is proposed for attaining 3D deployment and dynamic movement of multiple UAVs. Firstly, a genetic algorithm based K-means (GAK-means) algorithm is utilized for obtaining the cell partition of the users. Secondly, Q-learning based deployment algorithm is proposed, in which each UAV acts as an agent, making their own decision for attaining 3D position by learning from trial and mistake. In contrast to the conventional genetic algorithm based learning algorithms, the proposed algorithm is capable of training the direction selection strategy offline. Thirdly, Q-learning based movement algorithm is proposed in the scenario that the users are roaming. The proposed algorithm is capable of converging to an optimal state. Numerical results reveal that the proposed algorithms show a fast convergence rate after a small number of iterations. Additionally, the proposed Q-learning based deployment algorithm outperforms K-means algorithms and Iterative-GAKmean (IGK) algorithms with low complexity.
\end{abstract}


\section{Introduction}
The unprecedented demands for the high quality of wireless services impose enormous challenges for conventional cellular communication networks. UAV-assisted communications in which the UAVs are employed as aerial base stations for assisting the existing terrestrial communication infrastructure, has been viewed as a promising candidate technique for tackling with problems like quick-response wireless service recovery after unexpected infrastructure damage or natural disasters, as well as cellular network offloading in hot spots such as sport stadiums, outdoor events~\cite{kosmerl2014ICC,zeng2016COM,osseiran2014COM}. Additionally, UAVs can also be employed in the Internet of Things (IoT) networks that require massive connectivity~\cite{qin2018Sparse,qin2019low,zhang2019cellulartcom}. In IoT networks, UAVs are capable of collecting data from ground IoT devices in a given geographical area, in which constructing a complete cellular infrastructure is unaffordable~\cite{Mozaffari2017WCOM,wang2018joint}.

Given the advantages of agility and mobility, low cost, their beneficial line-of-sight (LoS) propagation, the application of UAV-assisted communication networks is highly desired due to the following advantages:

\begin{itemize}
\item \textbf{Agility feature:} UAV can be deployed quickly and easily to offer high-quality services in quick response scenarios. Additionally, the cost of deploying UAV as an aerial base station is much more affordable than terrestrial base stations~\cite{zeng2016COM}.
\item \textbf{LoS feature:} As is deployed in the three-dimensional space, UAV has the feature of LoS connections towards ground users which can be significantly less affected by shadowing and fading~\cite{khawaja2017uav}.
\item \textbf{Mobility feature:} UAV is capable of adjusting their 3D positions dynamically and swiftly according to users' real-time locations, which enables it to provide flexible and on-demand service for users~\cite{wang2018joint}.
\end{itemize}

Sparked by the aforementioned characteristics of UAV-assisted communication networks, American wireless carriers AT\&T~\cite{zhang2017Optimal} have conducted the research of employing UAVs to develop airborne LTE services. UAVs have been utilized by Huawei corporation and China mobile corporation as aerial base stations in emergency communications, and their achievements had been exhibited in 2018 Mobile World Congress (MWC) in Barcelona. Additionally, UAV-based aerial base stations which equipped with the fifth generation (5G) networks has experienced its first experiment in April 2018 by Huawei cooperation in Shenzhen. With the number and requirement of UAVs increasing more than tenfold year by year, it faces enormous challenges to designing UAV-assisted wireless networks.

\subsection{Related Works}

In order to fully reap the benefits of the deployment of UAVs for communication purposes, some core technical challenges need to be faced with firstly. For instance, the 3D placement of multiple UAVs~\cite{hu2018reinforcement,Mozaffari2019Beyond}, the interference elimination~\cite{van2016lte}, energy efficiency optimization~\cite{zeng2017WCOM,kandeepan2014aerial,li2016TMC}, trajectory/movement design~\cite{liu2018comp,zhang2018joint,fadlullah2016dynamic}, air-ground channel modeling~\cite{khawaja2017uav}, antenna design~\cite{Mozaffari2019Communications}, power/bandwidth allocation~\cite{wang2018joint,zhang2019cellular,takaishi2017virtual,tang2018ac} and the compatibility between aerial wireless networks and the existing cellular networks. Among all these challenges, the deployment of UAVs is the fundamental and foremost one. Some groups have made great contributions to extensively solving the aforementioned problems.

\subsubsection{Deployment of Static UAVs}

Early research contributions had studied the deployment of static UAVs for both single-UAV scenarios and multi-UAV scenarios. The authors in~\cite{kosmerl2014ICC} proposed a solution for the deployment of a single aerial base station in the scenario that ground base stations were destroyed after a natural disaster. In the letter~\cite{alhourani2014WCOML}, the authors proposed an analytical solution for optimizing the UAV's altitude for providing maximum radio coverage for the users. In~\cite{alzenad2017WCOML}, the authors presented an optimal deployment algorithm for UAV-BS to provide wireless service to the maximal number of connected ground users. At the meantime, the transmit power of UAV is also minimized. However, these research contributions mainly focus on the problem of a single UAV. The cooperative deployment of multiple-UAV needs to be designed for providing reliable service for users. In~\cite{Mozaffari2017IEEE_J_WCOM}, the authors proposed a novelty framework for maximizing the throughput of users while considering the fairness between the users. Three-dimensional placement of the UAVs was obtained with the aid of the circle packing theory. Both the total coverage radium and operating lifetime of the UAVs were maximized in~\cite{mozaffari2016COML}. However, the high mobility and dynamic movement design of multiple UAVs are not considered in the researches mentioned above.

\subsubsection{Deployment of mobile UAVs}

It is intuitionist that UAV with movements is capable of obtaining better performance than a static UAV in the scenario that users are mobile. When faced with the challenge of trajectory design, current researches mainly consider the scenario that ground users are static. The trajectory of a single UAV was designed in~\cite{Lyu2016Cyclical} for serving each user via TDMA, even though orthogonal multiple access~\cite{liu2017nonorthogonal} had been proposed to improve the system capacity for 5G networks. Finally, a cyclical trajectory was designed for providing more reliable services for users. Additionally, a significant throughput gain over the case of a single static UAV was obtained. In~\cite{wu2018common}, the authors presented a low-complexity solution for the trajectory design of a single UAV for maximizing the minimum average throughput of all users. However, only a single UAV was considered in this research, and the trajectory of UAV was designed as a simple circular trajectory. Research on the dynamic movement of multiple-UAV based on the mobility of users is still a blank domain.

\subsubsection{Reinforcement learning in UAV-assisted communications}

Machine learning applications have gained remarkable attention in wireless communication networks during recent decades~\cite{qin2019deep}, and the reinforcement learning algorithm has been proved to possess the capacity of tackling problems in UAV-assisted wireless networks~\cite{Galindo2010Distributed,wang2019autonomous,hu2018reinforcement,liu2019machine}. In order to support dynamic user grouping and bring more flexibility to network design, the authors in~\cite{Liu2019Communications} jointly designed the trajectory and power allocation of UAV for serving static non-orthogonal multiple access (NOMA) users. The design challenges of integrating NOMA techniques into UAV networks have been investigated while some open research issues are also highlighted. An interference-aware path planning scheme based on deep reinforcement learning was proposed for a network of cellular-connected UAVs in~\cite{challita2018cellular}, better wireless latency and transmit rate was achieved in the proposed scheme. In ~\cite{lu2018uav}, the authors proposed a UAV relay scheme based on both reinforcement learning and deep reinforcement learning. With the aid of these two algorithms, the energy consumption of the UAV is minimized while a better bit error rate (BER) performance is received. However, the multiple-UAV scenario was not considered in this article. In ~\cite{liu2018energy}, the authors invoked a deep reinforcement learning (DRL) algorithm for energy efficient control of UAVs by jointly considering communications coverage, fairness, energy consumption, and connectivity. The aim is to find a control policy that specifies how each UAV moves in each timeslot. Thus, four parameters: average coverage score, fairness index, average energy consumption and energy efficiency are jointly optimized. However, the movement of ground users was also neglected for the purpose of simplifying the system model.

\subsection{Our New Contributions}

Again, deploying UAVs as aerial base stations have seen increasing applications both in academia and industry~\cite{zhang2017cellular}. However, currently, few research contributions consider the 3D deployment of multi-UAV. The real-time movement of ground users is also overlooked by current articles, based on which, the mobility of UAVs needs to be considered to attain a better quality of service. In contrast to current articles~\cite{Lyu2016Cyclical,challita2018cellular} that design the trajectory of UAV with fixed source and destination, the movement of UAVs in this article has no fixed destination due to the mobility of users. Additionally, the research gap still exists on investigating the QoE-driven UAVs-assisted communications, which is also worth studying for further improving the wireless services. The motivations of this work are concluded as follows:

\begin{itemize}
\item Mobility of UAVs has not been considered based on the movement of users in most current research contributions, which mainly focus on the two-dimensional placement of multi-UAV or the mobility of single-UAV while ground users remain static. The UAVs need to travel in the scenario that ground users are roaming continuously to maintain good connections.
\item QoE has not been considered for UAVs-assisted communications in most current research contributions, which mainly address this issue without considering the specific requirements of different ground users. QoE is invoked for demonstrating the users' satisfaction, and it is supposed to be considered in UAV-assisted wireless networks.
\item Three-dimensional dynamic movement of UAVs has not been considered in most current research contributions, which mainly focus on the 2D movement design of UAVs. UAVs are capable of adjusting their 3D positions dynamically and swiftly to maintain a high performance. In this case, the 3D dynamic movement design of UAVs needs to be investigated.
\end{itemize}

Therefore, the problem of 3D deployment and dynamic movement design of multiple UAVs is formulated for improving the users' QoE instead of throughput. Our new contributions are concluded as follows:

\begin{enumerate}
\item We propose a novel QoE-driven multi-UAV assisted communication framework, in which multiple UAVs are deployed in a 3D space to serve mobile users. The mean opinion score (MOS) is adopted for evaluating the satisfaction of users. Meanwhile, we formulate the sum MOS maximization problem by optimizing the UAVs' placement and dynamic movement.
\item We develop a three-step approach for solving the formulated problem. i) We invoke the GAK-means algorithm for obtaining the initial cell partition. ii) We develop a Q-learning based deployment algorithm for attaining 3D placement of UAVs when users are static at the initial time. iii) We develop a Q-learning based movement algorithm for designing the 3D dynamic movement of the UAVs.
\item We conceive a Q-learning based solution to solve the NP-hard 3D deployment and movement problem of the UAVs. In contrast to conventional genetic-based learning algorithms, the proposed algorithm is capable of training the policy by learning from the trial and mistake.
\item We demonstrate that the proposed algorithms show a fast convergence. Additionally, the proposed Q-learning based deployment algorithm outperforms K-means algorithms and IGK algorithms with low complexity.
\end{enumerate}

\subsection{Organization}

In Section II, the problem formulation for the QoE-based 3D deployment and dynamic movement of multi-UAV is presented. In Section III, efficient cell partition and UAVs' deployment for QoE maximization with static users are proposed. In Section IV, the Q-learning algorithm is utilized for attaining the UAVs' dynamic movement when users are roaming. The numerical results for both deployment and dynamic movement of multi-UAV are illustrated in Section V. The conclusions are presented in Section VI, which is followed by some essential proofs. Additionally, The list of notations is demonstrated in Table~\ref{List of Notations}.

\begin{table*}[tbp]\footnotesize

\caption{List of Notations}
\centering
\begin{tabular}{|c||c|c||c|}
\hline
\centering
 Notations & Description & Notations & Description\\
\hline
\centering
 $U$ & Ground users' number & $N$ & Cluster number and UAVs' number\\
\hline
\centering
 ${x_{{k_n}}},{y_{{k_n}}}$ & Ground users' coordinates & ${x_n},{y_n}$ & UAVs' coordinate\\
\hline
\centering
 $f_c$ & Carrier frequency & $h_n$ & UAVs' altitudes\\
\hline
\centering
 ${P_{\max }}$ & UAV's maximum transmit power & ${g_{{k_n}}}$ & channel power gain from UAVs to users\\
\hline
\centering
 $N_0$ & Noise power spectral & $B$ & Total bandwidth \\
\hline
\centering
 ${\mu _{LoS}},{\mu _{NLoS}}$ & Additional path loss for LoS and NLoS & ${P_{LoS}},{P_{NLoS}}$ & LoS and NLoS probability\\
\hline
\centering
 ${\Gamma _{{k_n}}}$ & Receives SINR of ground users & ${I_{{k_n}}}$ & Receives interference of ground users\\
\hline
\centering
 ${r_{{k_n}}}$ & Instantaneous achievable rate & ${R_{sum}}$ & Overall achievable sum rate\\
\hline
\centering
 $b_1$,$b_2$ & Environmental parameters (dense urban) & $\alpha $ & Path loss exponent\\
\hline
\centering
 ${\text{MOS}}({k_n}) $ & Overall achievable MOS & $a_t$ & State in Q-learning model at time $t$\\
 \hline
\centering
 $a_t$ & Action in Q-learning model at time $t$ & $r_t$ & Reward in Q-learning model at time $t$\\
 \hline
\end{tabular}
\label{List of Notations}
\end{table*}

\section{System Model}

\subsection{System Description}
\begin{figure} [t!]
\centering
\includegraphics[width=3.2in]{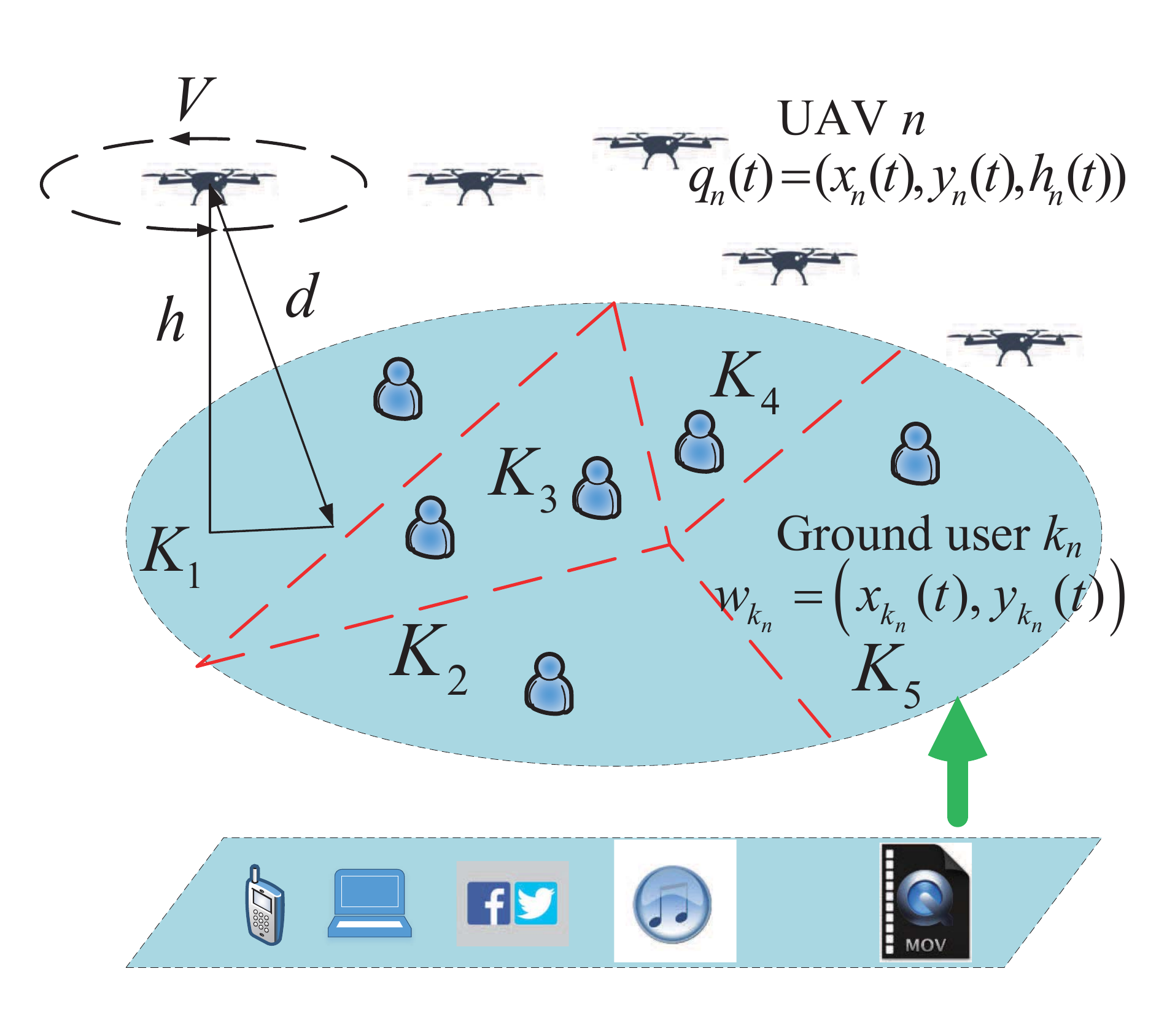}
\caption{QoE-driven deployment of multiple UAVs in wireless communications.}\label{qoe}
\end{figure}

Consider the down-link transmission in UAV-assisted wireless networks, where multi-UAV are deployed as aerial base stations for supporting the mobile users in a particular area. As illustrated in Fig.~\ref{qoe}, this particular area is divided into $N$ clusters, then the users are denoted as $K = \left\{ {{K_1}, \cdots {K_N}} \right\}$, while ${K_n}$ is the users partitioned into cluster $n$, $n \in \mathbb{N} = \left\{ {1,2, \cdots N} \right\}$. The clustering scheme will be discussed in Section III. Each user belongs to only one cluster, thus ${K_n} \cap {K_{{n'}}} = \phi ,{n'} \ne n,\forall {n'},n \in \mathbb{N}$, while ${{\rm K}_n} = \left| {{K_n}} \right|$ is the total number of users partitioned to the cluster $n$. In any cluster $n$, $n \in \mathbb{N}$, a single UAV is dispatched. The UAVs are able to connect with the core network by satellite or a fully working terrestrial base station. At any time $t$, multiple users in the same cluster are served by the same UAV simultaneously by employing FDMA~\cite{Mozaffari2016Unmanned}. The coordinate of each user is denoted by ${w_{{k_n}}} = {[{x_{{k_n}}}(t),{y_{{k_n}}}(t)]^T} \in {\mathbb{R}^{2 \times 1}},{k_n} \in {K_n}$, where ${x_{{k_n}}}(t)$ and ${y_{{k_n}}}(t)$ are the coordinates of user $k_n$ at time $t$. The movement of ground users is discussed in section IV. The vertical trajectory is denoted by ${h_n}(t) \in [{h_{\min }},{h_{\max }}],0 \le t \le {T_s}$. The horizontal coordinate of the UAV is denoted by ${q_n}(t) = {[{x_n}(t),{y_n}(t)]^T} \in {\mathbb{R}^{2 \times 1}}$,with $0 \le t \le {T_s}$\footnote[1]{In this paper, the velocity of UAVs is constant, the scenario that UAVs have changeable velocity will be discussed in our future work.}.

Furthermore, the distance from the UAV $n$ to user $k_n$ at time $t$ can be expressed as
\begin{align}\label{dt}
{{d_{{k_n}}}(t) = \sqrt {{h_n}^2(t) + {{[{x_n}(t) - {x_{{k_n}}}(t)]}^2} + {{[{y_n}(t) - {y_{{k_n}}}(t)]}^2}}}.
\end{align}
\subsection{Signal Model}

As are deployed in the 3D space, the UAVs have a higher probability of LoS connections than terrestrial communication infrastructures. The LoS probability is given by~\cite{alhourani2014WCOML}
\begin{align}\label{plos}
{{P_{{\text{LoS}}}}({\theta _{{k_n}}}) = {b_1}{(\frac{{180}}{\pi }{\theta _{{k_n}}} - \zeta )^{{b_2}}}},
\end{align}
where ${\theta _{{k_n}}}(t) = {\sin ^{ - 1}}[\frac{{{h_n}(t)}}{{{d_{{k_n}}}(t)}}]$ is the elevation angle between UAV $n$ and the ground user ${k_n}$. $b_1$, $b_2$ and $\zeta $ are constant values determined by environment. In the particular case, the altitude of the $n$-UAV ${h_n}(t) \in [{h_{\min }},{h_{\max }}],0 \le t \le {T_s}$ is supposed to be properly chosen in practice to balance between the probability of LoS channel and the path loss. Then, the Non-Line-of-Sight (NLoS) probability is given by ${P_{{\text{NLoS}}}} = 1 - {P_{{\text{LoS}}}}$.

Thus, the channel power gain from UAV $n$ to user $k_n$ at time $t$ can be expressed as
\begin{align}\label{gt}
{{g_{{k_n}}}(t) = {K_0}^{ - 1}d_{{k_n}}^{ - \alpha }(t){[{P_{{\text{LoS}}}}{\mu _{{\text{LoS}}}} + {P_{{\text{NLoS}}}}{\mu _{{\text{NLoS}}}}]^{ - 1}}},
\end{align}
where ${K_0} = {\left( {\frac{{4\pi {f_c}}}{c}} \right)^2}$, $\alpha $ is a constant denotes path loss exponent,  ${\mu _{LoS}}$ and ${\mu _{NLoS}}$ are different attenuation factors considered for LoS and NLoS links, $f_c$ is the carrier frequency, and finally $c$ is the speed of light.

Denote that the total available bandwidth $B_n$ of UAV $n$ is distributed to the associated user equally, thus the bandwidth of each user is expressed as ${B_{{k_n}}} = {B_n \mathord{\left/
 {\vphantom {B {{K_n}}}} \right.
 \kern-\nulldelimiterspace} {{K_n}}}$. The spectrum utilized by the different clusters is different in this paper, the case that allocating same spectrum to each cluster will be considered in our future work. Also denote that the total transmit power is allocated to each users uniformly, namely ${p_{{k_n}}} = {{{P_{\max }}} \mathord{\left/
 {\vphantom {{{P_{\max }}} {{K_n}}}} \right.
 \kern-\nulldelimiterspace} {{K_n}}}$, where ${P_{\max }}$ is the maximum allowed transmit power of each UAV. The transmit power from UAVs to the associated users is constant, the dynamic power allocation will be considered in our future work.

As the received interference from UAV to users is able to be mitigated by allocating different spectrum to each cluster\cite{liu2018comp,zhang2017Optimal}, the received SNR ${\Gamma _{{k_n}}}(t)$ of ground user $k_n$ connected to UAV $n$ at time $t$ can be expressed as
 \begin{align}\label{sinr}
{{\Gamma _{{k_n}}}(t) = \frac{{{p_{{k_n}}}{g_{{k_n}}}(t)}}{{  {\sigma ^2}}}},
\end{align}
where ${\sigma ^2} = {B_{{k_n}}}N_0^{}$ with $N_0$ denoting the power spectral density of the additive white Gaussian noise (AWGN) at the receivers. As the requirements of transmit rate for different users vary, the SNR target which must be received by each ground users is ${\gamma _{{k_n}}}$, then ${\Gamma _{{k_n}}}(t) \ge {\gamma _{{k_n}}} $.

\begin{lemma}\label{transmit power}
For the aim of guaranteeing that all the users are capable of connecting to the networks, we have a constraint for the UAVs' transmit power, which can be expressed as
\begin{align}\label{pmax}
{{P_{\max }} \geqslant \gamma {\sigma ^2}{K_0}d_{{k_n}}^\alpha (t){\mu _{NLoS}}}.
\end{align}

\begin{proof}
See Appendix A~.
\end{proof}
\end{lemma}
\vspace{0.3cm}

\textbf{Lemma~\ref{transmit power}} shows the constraint for transmitting power of UAVs. When choosing the type of UAVs in practical application, the lower bound of transmit power is supposed to be no less than equation (5). In this case, the received SNR of every ground users can be larger than the SNR threshold.

The user $k_n$'s transmit rate ${r_{{k_n}}}(t)$ at time $t$ can be expressed as
\begin{align}\label{rt}
{{r_{{k_n}}}(t) = {B_{{k_n}}}{\log _2}[1 + \frac{{{p_{{k_n}}}{g_{{k_n}}}(t)}}{{{\sigma ^2}}}]}.
\end{align}

\begin{proposition}\label{lower and upper}
The altitude constraint for the UAV $n$ has to satisfy
 \begin{align}\label{Rt}
{{d_{{k_n}}}(t)\sin \left[ {\frac{\pi }{{180}}(\zeta  + {e^{M(t)}})} \right] \le {h_n}(t) \le {\left( {\frac{{{P_{\max }}}}{{\gamma {K_0}{\sigma ^2}{\mu _{LoS}}}}} \right)^{{1 \mathord{\left/
 {\vphantom {1 \alpha }} \right.
 \kern-\nulldelimiterspace} \alpha }}}},
\end{align}
where $M(t) = {{\left[ {\ln ({{\frac{S(t)}{{({\mu _{LoS}} - {\mu _{NLoS}})}} - \frac{{{\mu _{NLoS}}}}{{{\mu _{LoS}} - {\mu _{NLoS}}}}} \mathord{\left/
 {\vphantom {{\frac{S}{{({\mu _{LoS}} - {\mu _{NLoS}})}} - \frac{{{\mu _{NLoS}}}}{{{\mu _{LoS}} - {\mu _{NLoS}}}}} {{b_1}}}} \right.
 \kern-\nulldelimiterspace} {{b_1}}})} \right]} \mathord{\left/
 {\vphantom {{\left[ {\ln ({{\frac{S(t)}{{({\mu _{LoS}} - {\mu _{NLoS}})}} - \frac{{{\mu _{NLoS}}}}{{{\mu _{LoS}} - {\mu _{NLoS}}}}} \mathord{\left/
 {\vphantom {{\frac{S(t)}{{({\mu _{LoS}} - {\mu _{NLoS}})}} - \frac{{{\mu _{NLoS}}}}{{{\mu _{LoS}} - {\mu _{NLoS}}}}} {{b_1}}}} \right.
 \kern-\nulldelimiterspace} {{b_1}}})} \right]} {{b_2}}}} \right.
 \kern-\nulldelimiterspace} {{b_2}}}$ and $S(t) = \frac{{{P_{\max }}}}{{\gamma {K_0}{\sigma ^2}d_{{k_n}}^\alpha (t)}}$. Other parameters are introduced in Section II.

\begin{proof}
See Appendix B~.
\end{proof}
\end{proposition}
\vspace{0.3cm}

\textbf{Proposition~\ref{lower and upper}} provides the necessary conditions for the UAV's altitude needed to be capable of providing reliable services for associated users. From Proposition, it is observed that the lower bound of the altitude is the function of distance ${d_{{k_n}}}(t)$. At the same time, the upper bound of the altitude is the function of maximum transmit power ${P_{\max }}$. In this case, as the distance and transmit power vary, the altitude of the corresponding UAVs needs to be adjusted for proving reliable services to users.

\subsection{Quality-of-Experience Model}

As illustrated in Fig.1, the requirements of the transmit rate for different users vary. In this case, the QoE model is essential to be considered in UAV-assisted communication networks.

\begin{definition}\label{difinition:qoe}
QoE is a metric that depends on a person's preferences towards a particular object or service related to expectations, experiences, behavior, cognitive abilities, object's attributes, and the environment~\cite{gao2018qoe}.
\end{definition}

MOS is invoked as the measure of the user's QoE~\cite{Mitra2015QoE,Cui2017QoE}, and the MOS model is defined as follow
\begin{align}\label{mos1e}
{{\text{MO}}{{\text{S}}_{{k_n}}}(t) = {\xi _1}{\text{MOS}}_{{k_n}}^{{\text{delay}}}(t) + {\xi _2}{\text{MOS}}_{{k_n}}^{{\text{rate}}}(t)},
\end{align}
where ${\xi _1}$ and ${\xi _2}$ are coefficients and ${\xi _1} + {\xi _2} = 1$.

\begin{definition}\label{difinition:mos}
We consider five hypotheses in relation to the five alternatives (on the ordinal scale) for each QoE state, the measurement is the MOS received by users. These are: excellent (4.5), very good ($3.5 \sim 4.5$), good ($2 \sim 3.5$), fair ($1 \sim 2$) and poor (1).
\end{definition}

Following ~\cite{Cui2017QoE,rugelj2014novel}, we consider web browsing applications in this paper, in which ${\text{MOS}}_{{k_n}}^{{\text{delay}}}(t)$ is ignored. Thus, the MOS model is simplified as follows~\cite{rugelj2014novel}
 \begin{align}\label{mos22}
{{\text{MO}}{{\text{S}}_{{k_n}}}(t) =  - {C_1}\ln [d({r_{{k_n}}}(t))] + {C_2}},
\end{align}
where $d\left( {{r_{{k_n}}}(t)} \right)$ is the delay time related to the transmit rate, while ${\text{MO}}{{\text{S}}_{{k_n}}}(t)$ denotes the MOS score ranging from 1 to 4.5. Finally, $C_1$ and $C_2$ are constants determined by analyzing the experimental results of the web browsing applications, which are set to be 1.120 and 4.6746, respectively~\cite{rugelj2014novel}.

The delay time in (9) is given by~\cite{rugelj2014novel}
\begin{align}\label{delay}
{\begin{gathered}
  d({r_{{k_n}}}(t)) = 3{\text{RTT}} + \frac{{{\text{FS}}}}{{{r_{{k_n}}}(t)}} + L\left( {\frac{{{\text{MSS}}}}{{{r_{{k_n}}}}}} \right) \hfill \\
  {\kern 1pt} {\kern 1pt} {\kern 1pt} {\kern 1pt} {\kern 1pt} {\kern 1pt} {\kern 1pt} {\kern 1pt} {\kern 1pt} {\kern 1pt} {\kern 1pt} {\kern 1pt} {\kern 1pt} {\kern 1pt} {\kern 1pt} {\kern 1pt} {\kern 1pt} {\kern 1pt} {\kern 1pt} {\kern 1pt} {\kern 1pt} {\kern 1pt} {\kern 1pt} {\kern 1pt} {\kern 1pt} {\kern 1pt} {\kern 1pt} {\kern 1pt} {\kern 1pt} {\kern 1pt} {\kern 1pt}  + {\text{RTT}} - \frac{{2{\text{MSS}}\left( {{2^L} - 1} \right)}}{{{r_{{k_n}}}(t)}} \hfill \\
\end{gathered}  },
\end{align}
where RTT[s] is the round trip time, while FS[bit] is the web page size and MSS[bit] is the maximum segment size. The parameter $L = \min \left[ {{L_1},{L_2}} \right]$ represents the number of slow start cycles with idle periods, which are defined as ${L_1} = {\log _2}\left( {\frac{{{r_{{k_n}}}{\text{RTT}}}}{{{\text{MSS}}}} + 1} \right) - 1$ and ${L_2} = {\log _2}\left( {\frac{{{\text{FS}}}}{{{\text{2MSS}}}} + 1} \right) - 1$.

The sum MOS of user $k_n$ during a period $T_s$ can be expressed as
 \begin{align}\label{summos}
{{\text{MO}}{{\text{S}}_{{k_n}}} = \sum\limits_{t = 0}^{{T_s}} {{\text{MO}}{{\text{S}}_{{k_n}}}(t)}  }.
\end{align}

\begin{remark}\label{remark:qoe}
When the UAVs fly towards some users to acquire a better channel environment for these users, they get farther away from other users. Finally, the MOS of the users which are farther away from the UAVs become lower than those who are closer to the UAVs.
\end{remark}

Our approach can accommodate other MOS models without loss of generality. Meanwhile, the model invoked in this paper can also be extended to other applications in particular scenarios.

\subsection{Problem Formulation}

Let $Q = \left\{ {{q_n}(t),0 \le t \le {T_s}} \right\}$ and $H = \left\{ {{h_n}(t),0 \le t \le {T_s}} \right\}$. Again, we aim for optimizing the positions of the UAVs at each timeslots, i.e., $\left\{ {{x_n}(t),{y_n}(t),{h_n}(t)} \right\},{\kern 1pt} {\kern 1pt} {\kern 1pt} n = 1,2, \cdots N,{\kern 1pt} {\kern 1pt} {\kern 1pt} t = 0,1, \cdots {T_s}$, for maximizing the MOS of users. Our optimization problem is then formulated as

\begin{center}
\begin{subequations}\label{optimizationproblem}
\begin{align}
\mathop {\mathop {\max }\limits_{C,Q,H} {\kern 1pt} {\kern 1pt} {\kern 1pt} {\text{MO}}{{\text{S}}_{{\text{total}}}} = \sum\limits_{n = 1}^N {\sum\limits_{{k_n} = 1}^{{{\text{K}}_n}} {\sum\limits_{t = 0}^{{T_s}} {{\text{MO}}{{\text{S}}_{{k_n}}}(t)} } } } ,\\
{\text{s}}{\text{.t}}{\text{.}}{\kern 1pt} {\kern 1pt} {\kern 1pt} {\kern 1pt}{\kern 1pt} {\kern 1pt} {\kern 1pt} {\kern 1pt} {\kern 1pt} {\kern 1pt} {\kern 1pt} {\kern 1pt} {\kern 1pt} {\kern 1pt} {K_n} \cap {K_{{n'}}} = \phi ,{n'} \ne n,\forall n,  \\
   {h_{\min }} \le {h_n}(t) \le {h_{\max }}, \forall t,\forall n,\\
   {\Gamma _{{k_n}}}(t) \ge \gamma_{k_n} ,\forall t,\forall {k_n},\hfill \\
   \sum\limits_{{k_n} = 1}^{{{\rm K}_n}} {{p_{{k_n}}}(t)}  \le {P_{\max }},\forall t,\forall {k_n},  \\
   {p_{{k_n}}}(t) \ge 0,\forall {k_n},\forall t,
\end{align}
\end{subequations}
\end{center}
where ${{{\text{K}}_n}}$ is the number of users partitioned to the cluster $n$, ${h_n}(t)$ is the altitude of UAV $n$, while ${p_{{k_n}}}(t)$ is the transmit power form UAV $n$ to ground user $k_n$. Furthermore, (12b) denotes that each user is partitioned into a single cluster; (12c) indicates the altitude constraint of the UAVs; (12d) formulates the minimized SINR target for each user to connect to the network; (12e) and (12f) qualifies the transmit power constraint of UAVs. Given the bandwidth and power allocation of the UAVs, this problem is then simplified into the problem of UAVs' movement optimization. Since the movement of users affect the received transmit rate to satisfy their QoE requirement, the UAVs have to travel based on the real-time movement of users to maintain the QoE of users, the problem of maximizing the MOS of the UAVs in (12a) inherently incorporates the design of dynamic movement for the UAVs.

\begin{theorem}\label{non-convex}
Problem (12a) is a non-convex problem since the objective function is non-convex over ${x_n}(t)$'s, ${y_n}(t)$'s and ${h_n}(t)$'s~\cite{zeng2017WCOM,wang2018joint}.
\begin{proof}
See Appendix C~.
\end{proof}
\end{theorem}

From equation (12a), it can be observed that the MOS of users is impacted by the UAVs' altitude. This is because both the distance and LoS probability between the UAVs and the users are related to the UAVs' altitudes. Increasing the UAV's altitude leads to a higher path loss while a higher LoS probability is obtained, the UAVs have to increase the transmit power for satisfying the users' QoE requirements.

\begin{remark}\label{remark:mos}
The sum MOS depends on the transmit power, number and positions (both horizontal positions and altitudes) of the UAVs.
\end{remark}

\begin{figure} [t!]
\centering
\includegraphics[width=3.2in]{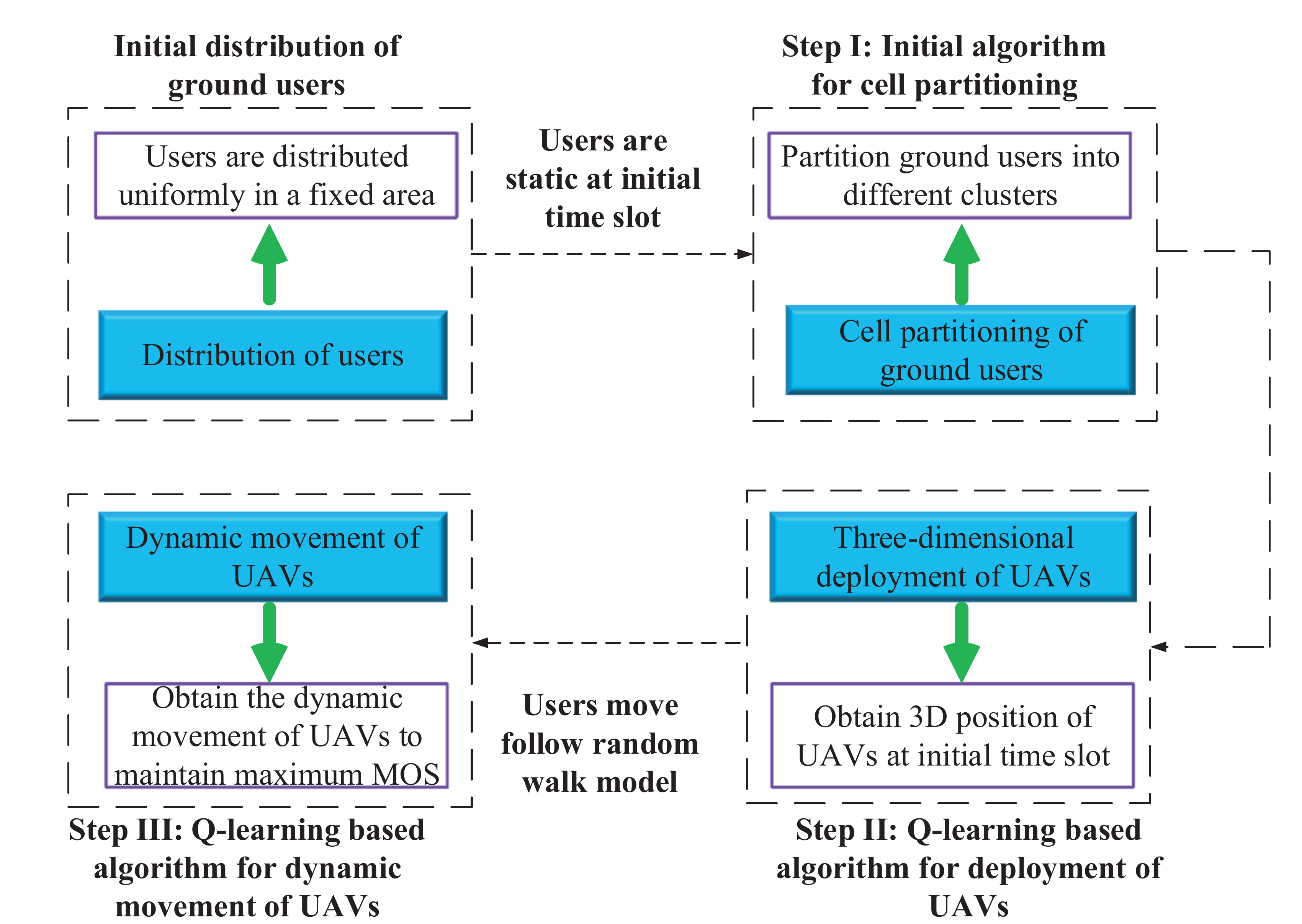}
 \caption{The procedure for solving the problem of deploying and movement design of multiple UAVs.
 }\label{suanfa}
\end{figure}


Due to the combinatorial features of cell partition and optimal position searching, an exhaustive search solution provides a straightforward method for obtaining the final results. We propose an efficient GAK-means based IGK algorithm for solving this problem. In addition, a low-complexity Q-learning based 3D deployment algorithm is also developed for the aim of avoiding the IGK algorithm's huge complexity. Furthermore, when the initial position of UAVs is fixed, obtaining the dynamic movement is also nontrivial due to the non-convex property of the problem (12) in terms of the sum MOS maximization. Q-learning algorithm is proved to be an efficient solution for tackling the problem which has exact transition formulation. This is the motivation of applying the Q-learning algorithm to tackle with this problem. This scheme is discussed in the next section\footnote[2]{In our future work, we will consider the online design of UAVs' trajectories, and the mobility of UAVs will be constrained to 360 degrees of angles instead of 7 directions. Given that, the state-action space is huge, a deep multi-agent Q-network based algorithm will be proposed in our future work.}.

\section{Three-Dimensional deployment of the UAVs}

We consider the scenario that the UAV $n$ is hovering above users with an alterable altitude while the users are static. The bandwidth and the transmit power of each UAV are allocated uniformly to each user. Thus, the optimization problem is simplified to a region segmentation problem, which is formulated as
\begin{subequations}\label{optimizationproblem2}
\begin{align}
\mathop {\mathop {\max }\limits_{C{\text{,}}Q,H} {\kern 1pt} {\kern 1pt} {\kern 1pt} {\text{ MO}}{{\text{S}}_{{\text{total}}}} = \sum\limits_{n = 1}^N {\sum\limits_{{k_n} = 1}^{{{\rm K}_n}} {{\text{MO}}{{\text{S}}_{k_n}}} }} ,\\
{\text{s}}{\text{.t}}{\text{.}}{\kern 1pt} {\kern 1pt} {\kern 1pt} {\kern 1pt}{\kern 1pt} {\kern 1pt} {\kern 1pt} {\kern 1pt} {\kern 1pt} {\kern 1pt} {\kern 1pt} {\kern 1pt} {\kern 1pt} {\kern 1pt}{K_n} \cap {K_{{n'}}} = \phi ,{n'} \ne n,\forall n,  \\
   {h_{\min }} \le {h_n}\le {h_{\max }}, \forall n, \\
   {\Gamma _{{k_n}}} \ge \gamma_{K_n}  , \forall {k_n} \\
   \sum\limits_{{k_n} = 1}^{{{\rm K}_n}} {{p_{{k_n}}}}  \le {P_{\max }}, \forall {k_n}\\
   {p_{{k_n}}}\ge 0,\forall {k_n}.
\end{align}
\end{subequations}

\begin{theorem}\label{NP-hard}
Problem of (13a) is a NP-hard problem. More specifically, problem (13a) is NP-hard even only consider the users clustering.

\begin{proof}
See Appendix D~.
\end{proof}
\end{theorem}

\subsection{Initial Algorithm for Cell Partition of Ground Users}

User clustering is the first step for obtaining the deployment of the UAVs. It's the process of partitioning the users into different clusters. In each cluster, the users are linked with a UAV. In this model, users need to be partitioned into different clusters. In each cluster, a single UAV is employed. K-means algorithm, also named Lloyd algorithm can solve the problem of clustering and obtaining the initial 3D position of the UAVs with low complexity. This algorithm is capable of partitioning users into different clusters based on the policy of nearest neighbor barycenter and recalculate the barycenter of each cluster~\cite{kanungo2002efficient}. Specifically, the K-means algorithm is capable of obtaining the results that the squared error between the empirical mean of a cluster and the points is minimized~\cite{cui2018application}. By invoking the K-means algorithm, the users are partitioned into $N$ clusters.

K-means, which is a greedy algorithm, can only converge to a local minimum, even though recent study has shown with a large probability K-means could converge to the global optimum when clusters are well separated K-means starts with an initial partition with $n$ clusters and assign patterns to clusters so as to reduce the squared error. Since the squared error always decreases with an increase in the number of clusters $n$, it can be minimized only for a fixed number of clusters.

As the K-means algorithm is very sensitive to the initial center, the performance of the K-means algorithm is deterministic over runs~\cite{kanungo2002efficient}. For the aim of supplying a gap of the K-means algorithm, GAK-means algorithm based on Genetic algorithm will be invoked for obtaining global optimum clustering results. Firstly, initialize the population ($N$ users) and find the best individuals as the center of cluster ${C_1}, \cdots, {C_N}$ based on the genetic algorithm. Secondly, deploy $N$ UAVs in each center ${\mu _n},{\kern 1pt} {\kern 1pt} {\kern 1pt} n = 1, \cdots ,N$, and compare the Euclid distance $\left( {{x_i},{\mu _n}} \right),{\kern 1pt} {\kern 1pt} {\kern 1pt} i = 1, \cdots ,U$, then partition user $i$ into the cluster with the smallest distance. Repeat this step until all users have been allocated, after which recompute the center of each cluster. Update the cluster member by repeating this step until the cluster members no longer change significantly.

\subsection{Q-learning Algorithm for 3D Deployment of UAVs}

In this section, given the cell partition of the users, our goal is for obtaining the 3D locations of the UAVs to maximize the sum MOS. In the process of cell partition, the GAK-means algorithm is capable of obtaining a group of locations of the UAVs which have the minimal Euclidean distance. As the MOS of a particular user is related to the distance between this user and the UAV, so GAK-means could be regarded as a low-complexity scheme to obtain deployment of the UAVs. However, the objective of GAK-means is obtaining minimal sum Euclidean distance, it can be noticed from (11) that the transmit rate is not only related to Euclidean distance between users and UAV but also related to the probability of LoS. So GAK-means is not effective to solve the 3D deployment problem. Although Iterative-GAKmean (IGK) algorithm is able to obtain the optimal altitudes of the UAVs through an iterative process, the complexity of the IGK algorithm is high, the worst case computational complexity of this algorithms is $O(N({U^2} + U + 1))$. In this case, a more effective algorithm is required to solve the 3D deployment problem.

In the Q-learning model, the UAVs act as agents, and the Q-learning model consists of four parts: states, actions, rewards, and Q-value. The aim of Q-learning is for attaining a policy that maximizes the observed rewards over the interaction time of the agents. At each time slot $t$ during iteration, each agent observes a state, ${s_t}$, from the state space, $S$. Accordingly, each agent carries out an action, ${a_t}$, from the action space, $A$, selecting the directions based on policy, $J$. The decision policy $J$ is determined by a Q-table, $Q({s_t},{a_t})$. The principle of policy at each time slot is to choose an action that makes the model obtain maximum Q-value. Following the action, the state of each agent transitions to a new state ${s_{t + 1}}$ and the agent receives a reward, ${r_t}$, determined by the instantaneous sum MOS of ground users.

\subsubsection{State Representation}

The individual agents are presented using a three-state UAV model defined as: $\xi  = (x_{{\text{UAV}}},y_{{\text{UAV}}},h_{{\text{UAV}}})$, where $(x_{{\text{UAV}}},y_{{\text{UAV}}})$ is the horizonal position of UAV and $h_{{\text{UAV}}}$ is the altitude of UAV. As the UAVs are deployed in a particular district, the state space can be denoted as ${x_{{\rm{UAV}}}}:\left\{ {0,1, \cdots {X_d}} \right\}$, ${y_{{\rm{UAV}}}}:\left\{ {0,1, \cdots {Y_d}} \right\}$, ${h_{{\rm{UAV}}}}:\left\{ {{h_{\min }}, \cdots {h_{\max }}} \right\}$, where ${X_d}$ and ${Y_d}$ are the maximum coordinate of this particular district.

For each trial, the initial state (position) of each UAV is determined randomly, the update rate of Q-table is determined by the initial position of the UAVs and the number of users and UAVs. The closer each UAV is placed with the optimal position; the faster the update rate will be.

\begin{algorithm}[!t]
\caption{Q-learning algorithm for 3-D deployment of UAVs}
\label{q1}
\begin{algorithmic}
\REQUIRE ~~\\
Q-learning structure, environment simulator.\\

\STATE  Cell partition of ground users based on K-means algorithm.
\STATE  Initialize $Q(s,a)$ arbitrarily, deploy UAVs at random points.
\REPEAT
\STATE  \textbf{  }\textbf{  }\textbf{  }\textbf{  }for each step of episode:\\
\STATE  \textbf{  }\textbf{  }\textbf{  }\textbf{  }Choose action $a$ according to policy derived from $Q$ to obtain maximum $Q$;
\STATE \textbf{  }\textbf{  }\textbf{  }\textbf{  } Carry out action $a$, observe reward $r$, update state ${s'}$;
\STATE  \textbf{  }\textbf{  }\textbf{  }\textbf{  }$Q(s,a) \leftarrow Q(s,a) + \alpha [r + \beta {\max _{a'}}Q(s',a') - Q(s,a)]$;
\STATE  \textbf{  }\textbf{  }\textbf{  }\textbf{  }$s \leftarrow {s'}$;
\STATE  \textbf{  }\textbf{  }\textbf{  }\textbf{  }Update Q-table based on reward $r$ to obtain maximum $r$;
\UNTIL{$s$ is terminal}
\ENSURE   Evaluation results.
\end{algorithmic}
\end{algorithm}

\subsubsection{Action Space}

At each step, each agent (UAV) carries out an action ${a_t} \in A$, which includes choosing a direction for themselves, according to the current state, ${s_t} \in S$, based on the decision policy $J$. Each UAV can fly towards a variety of directions which makes the problem non-trivial to solve. However, by assuming that the UAVs fly with simple coordinated turns, we can simplify the model down to 7 directions. We can obtain a tradeoff between the accuracy and model complexity when the directions are appropriately chosen.

\subsubsection{State Transition Model}

The transition from ${s_t}$ to ${s_{t + 1}}$ with reward ${r_t}$ when action ${a_t}$ is carried out can be characterized by the conditional transition probability $p({s_{t + 1}},{r_t}|{s_t},{a_t})$. The goal of the Q-learning model is for maximizing the long-term rewards expressed as follow
\begin{align}\label{longreward}
{{G_t} = {\rm E}[\sum\limits_{n = 0}^\infty  {{\beta ^n}{r_{t + n}}} ]},
\end{align}
where $\beta $ is the discount factor.

As shown in equation\eqref{longreward}, the goal of the Q-learning model is for maximizing the long term rewards. we conceive a three-step approach to solve the formulated problem. After obtaining the initial cell partition of users, in the second step, we invoke the Q-learning algorithm to find the optimal deployment of UAVs at an initial time regardless of the initial position of UAVs. In the third step, we aim at maximizing the total MOS of users during a period of $T_s$ by obtaining the movement of UAVs. These two steps are both long-term targets. As for the second step, we want to maximize the long-term rewards during the process of UAVs move from initial position to optimal position at the initial time. In terms of the third step, we want to maximize long-term rewards during a period of $T_s$.

The agents, states, actions, and rewards in Q-learning model are defined as follows:
\begin{itemize}
\item Agent: UAV $n$, $n \in \mathbb{N} = \left\{ {1,2, \cdots N} \right\}$.
\item State: The UAVs' 3D positions (horizontal coordinates and altitudes) are modeled as states.
\item Actions: Actions for each agent are the set of coordinates to present flying directions of the UAVs. (1,0,0) denotes the right turn of the UAV; (-1,0,0) indicates the left turn of the UAV; (0,1,0) represents the forward of the UAV; (0,-1,0) implies the backward of the UAV; (0,0,1) means the ascent of the UAV; (0,0,-1) indicates the descending of the UAV; (0,0,0) represents the UAV will stay static\footnote[3]{In this paper, the proposed algorithm can accommodate an arbitrary number of directions without loss of generality. We choose seven directions to strike a tradeoff between the performance and complexity of the system.}.
\item Rewards: The reward function is formulated by the instantaneous MOS of the users. If the action that the agent carries out at current time $t$ can improve the sum MOS, then the UAV receives a positive reward. The agent receives a negative reward if otherwise. Therefore, the reward function can be express as
   \begin{align}\label{reward}
{ {r_t} = \left\{ {\begin{array}{*{20}{c}}
  {{\kern 1pt} {\kern 1pt} {\kern 1pt} {\kern 1pt} {\kern 1pt} {\kern 1pt} {\kern 1pt} {\kern 1pt} {\kern 1pt} {\kern 1pt} {\kern 1pt} {\kern 1pt} 1,{\kern 1pt} {\kern 1pt} {\kern 1pt} {\kern 1pt} {\kern 1pt} {\kern 1pt} {\kern 1pt} {\kern 1pt} {\kern 1pt} {\kern 1pt} {\kern 1pt} {\kern 1pt} {\kern 1pt} {\kern 1pt} {\kern 1pt} {\kern 1pt} {\kern 1pt} {\kern 1pt} {\kern 1pt} {\kern 1pt} {\kern 1pt} {\kern 1pt} {\kern 1pt} {\kern 1pt} {\kern 1pt} {\kern 1pt} {\kern 1pt} {\kern 1pt} {\text{if}}{\kern 1pt} {\kern 1pt} {\text{MO}}{{\text{S}}_{{\text{new}}}}{\kern 1pt}  > {\text{MO}}{{\text{S}}_{{\text{old}}}},{\kern 1pt} } \\
  { - 0.1,{\kern 1pt} {\kern 1pt} {\kern 1pt} {\kern 1pt} {\kern 1pt} {\kern 1pt} {\kern 1pt} {\kern 1pt} {\kern 1pt} {\kern 1pt} {\kern 1pt} {\kern 1pt} {\kern 1pt} {\kern 1pt} {\kern 1pt} {\kern 1pt} {\kern 1pt} {\kern 1pt} {\kern 1pt} {\kern 1pt} {\kern 1pt} {\kern 1pt} {\kern 1pt} {\kern 1pt} {\text{if}}{\kern 1pt} {\kern 1pt} {\text{MO}}{{\text{S}}_{{\text{new}}}}{\kern 1pt} {\text{ = MO}}{{\text{S}}_{{\text{old}}}},} \\
  {{\kern 1pt} {\kern 1pt} {\kern 1pt} {\kern 1pt}  - 1,{\kern 1pt} {\kern 1pt} {\kern 1pt} {\kern 1pt} {\kern 1pt} {\kern 1pt} {\kern 1pt} {\kern 1pt} {\kern 1pt} {\kern 1pt} {\kern 1pt} {\kern 1pt} {\kern 1pt} {\kern 1pt} {\kern 1pt} {\kern 1pt} {\kern 1pt} {\kern 1pt} {\kern 1pt} {\kern 1pt} {\kern 1pt} {\kern 1pt} {\kern 1pt} {\kern 1pt} {\kern 1pt} {\kern 1pt} {\kern 1pt} {\text{if}}{\kern 1pt} {\kern 1pt} {\text{MO}}{{\text{S}}_{{\text{new}}}}{\kern 1pt}  < {\text{MO}}{{\text{S}}_{{\text{old}}}}.}
\end{array}} \right.}
\end{align}

\end{itemize}

\begin{remark}\label{remark:rreward}
Altering the value of reward does not change the final result of the algorithm, but its convergence rate is indeed influenced~\cite{matignon2006reward}.
\end{remark}

During learning, the state-action value function for each agent can be iteratively updated and it can be given as \eqref{qfunction} at the top of the next page.

\begin{figure*}
\begin{align}\label{qfunction}
{{Q_{t + 1}}({s_t},{a_t}) \leftarrow (1 - \alpha ) \cdot {Q_t}({s_t},{a_t}){\text{ +  }}\alpha  \cdot \left[ {{r_t} + \beta  \cdot {{\max }_a}{Q_t}({s_{t + 1}},a)} \right].}
\end{align}
\end{figure*}


The policy in Q-learning based deployment algorithm is chosen according to the $\epsilon $-greedy policy. More specifically, the policy $J$ that maximizes the Q-value is chosen with a high probability of $1-\epsilon $, while other actions are selected with a low probability to avoid staying in a local maximum, i.e.,

\begin{align}\label{sq7}
{\begin{gathered}
  Pr(J=\widehat{J}) =\left\{\begin{matrix}
1-\epsilon,  & \widehat{a}=\text{argmax}Q\left ( s,a \right ),
\\
\epsilon /\left ( \left | a \right |-1 \right ),& otherwise.
\end{matrix}\right.
\end{gathered}}
\end{align}

\begin{remark}\label{remark:diversity}
The convergence rate of the proposed algorithm for solving the problem of 3D deployment of the UAVs varies as the initial position of the UAVs is changed.
\end{remark}

\subsection{Analysis of the Proposed Algorithms}

\begin{table*}[htbp]\footnotesize
\caption{Complexity of four algorithms}
\begin{center}
\centering
\begin{tabular}{|c||c||c||c||c|}
\hline
\centering
Algorithm & K-means algorithm & IGK algorithm & Exhaust search & Q-learning algorithm\\
\hline
\centering
 Complexity & $O(TUN)$ & $O(TN({U^2} + U + 1))$ & $O(TU{N^U})$ & $O(TUN)$\\
\hline
\centering
Optimality & Suboptimal & Suboptimal & Optimal & Suboptimal\\
\hline
\centering
Stability & Not Stable & Not Stable & Stable & Stable\\
\hline
\end{tabular}
\end{center}
\label{table:Complexity}
\end{table*}

\subsubsection{Complexity} The computational complexity of the evaluation of square-error of a given solution string is $O(TU)$, the complexity of the mutation operator is $O(TU^2)$ and K-means operator is $O(TUN)$. Thus, the total complexity of GAK-means is calculated as $O(TN({U^2} + U + 1))$. As for the computational complexity of the Q-learning algorithm, through the training process off-line, when Q-table is updated, the complexity of the Q-learning algorithm and K-means algorithm is the same, which can be calculated as $O(TUN)$.

\subsubsection{Stability and Convergence} As the K-means algorithm is very sensitive to the initial center, the performance of the K-means algorithm and IGK algorithm is deterministic overruns. On the contrary, it has been proved in~\cite{Asheralieva2016An} that when learning rate $\alpha$ satisfies the conditions of convergence theorem $\sum\nolimits_{t \ge 0} {{\alpha _t} =  + \infty ,} $ and $\sum\nolimits_{t \ge 0} {\alpha _t^2}  <  + \infty $, regardless the initial settings of $Q(s,a)$, if trial time is sufficiently large, the Q-learning algorithm is capable of converging to the optimal value with probability 1. The core point of the Q-learning algorithm is updating Q-table. Once the Q-table is updated, the position and altitude of UAVs are founded, and the stability of the algorithm is stable.

\section{Dynamic movement design of UAVs}

In this section, a Q-learning based algorithm is proposed for obtaining UAVs' dynamic movement, which can be proved as non-convex~\cite{wang2018joint}. As the users are roaming at each time slot, the optimal position of UAVs in each cluster changes when the positions of the users change~\cite{liu2018comp}. There is no standard solution for solving the calculated problem proposed efficiently, for the aim of tackling with the challenges, the reinforcement learning approach will be invoked in this section.

\begin{remark}\label{remark:move}
As the users move continuously, the optimal positions of the UAVs change at each timeslot. The UAVs have to travel to offer a better downlink service. Otherwise, the users' QoE requirements will not be satisfied.
\end{remark}

Please note that, as UAVs and ground users move continuously, users may move away from their initial areas and mix with other users initially allocated to other UAVs. In this case, updating the users-UAV association according to the new location of both users and UAVs is capable of further enhancing the quality of service for ground users. Nevertheless, it requires raising the size of the action space. When ignoring the re-clustering, as we have constrained the mobility of UAVs to as few as 7 directions, the size of the action space is 7. When considering the re-clustering, the action space consists of two parts: selecting the flying directions and selecting the associated users. As $N$ is the number of UAVs and $\left| {{K_n}} \right|$ denotes the total number of users in the $n$-th cluster, the number of actions for user association is $2N\sum\nolimits_{n = 1}^N {\left| {{K_n}} \right|}  $. Thus the total size of action space is $7+2N\sum\nolimits_{n = 1}^N {\left| {{K_n}} \right|}  $. In this case, the complexity of the algorithm will increase tremendously, and it will be non-trivial to obtain a policy that maximizes the observed rewards because of a huge Q-table. The potential solution is invoking a deep reinforcement learning algorithm to reduce the dimension of Q-table, which will be discussed in our future work. In this paper, we study the trajectory design of UAVs during a short time. So, we assume that the user could not roam into other clusters. The practical mobility of users will be considered in our future work.

\begin{algorithm}[!t]
\caption{Training Stage Procedure for 3D dynamic movement of UAVs}
\label{Training}
\begin{algorithmic}[1]
\REQUIRE ~~\\
Initial position of UAVs derive from \textbf{Algorithm 1}, real-time movement of ground users.\\

\STATE  Deploy UAVs at the initial positions, random initialize the policy $J $.
\STATE  Repeat (for each step of episode):
\STATE  \textbf{  }\textbf{  }\textbf{  }\textbf{  }\textbf{  }Random choose an action $a$;
\STATE  \textbf{  }\textbf{  }\textbf{  }\textbf{  }\textbf{  }Carry out action $a$, observe reward $r$, update state ${s'}$;
\STATE  \textbf{  }\textbf{  }\textbf{  }\textbf{  }\textbf{  }Update the position of ground users;
\STATE  \textbf{  }\textbf{  }\textbf{  }\textbf{  }\textbf{  }Choose action $a$ according to policy derived from $Q$ to obtain maximum $Q$;
\STATE  \textbf{  }\textbf{  }\textbf{  }\textbf{  }\textbf{  }$s \leftarrow {s'}$;
\STATE \textbf{  }\textbf{  }\textbf{  }\textbf{  }\textbf{  }\textbf{  } Update Q-table based on reward $r$ to obtain maximum $r$.
\ENSURE  dynamic movement with maximum $r$.
\end{algorithmic}
\end{algorithm}

\begin{algorithm}[!t]
\caption{Testing Stage Procedure for 3D dynamic movement of UAVs}
\label{Testing}
\begin{algorithmic}[1]
\REQUIRE ~~\\
Initial position of UAVs derive from \textbf{Algorithm 1}, real-time movement of ground users.\\
\STATE  Deploy UAVs at the initial positions.
\STATE  Repeat (for each time slot):
\STATE  \textbf{  }\textbf{  }\textbf{  }\textbf{  }\textbf{  }\textbf{  }Select the action $a$ to obtain $\max Q$;
\STATE  \textbf{  }\textbf{  }\textbf{  }\textbf{  }\textbf{  }\textbf{  }Update the position of ground users.
\STATE  until terminal.
\ENSURE  3D dynamic movement of UAVs.
\end{algorithmic}
\end{algorithm}

Before obtaining the dynamic movement of UAVs, the mobility of the users is supposed to be considered firstly. There exists a variety of mobility models such as a deterministic approach, a hybrid approach, and a random walk model~\cite{Ren2017Acess,mingzhe2017JSAC}. In this paper, we choose the random walk model, also denoted as Markovian mobility model\footnote[4]{Our approach is capable of accommodating other mobility models without loss of generality.}, which can present the movement of the users. The direction of each user's movement is uniformly distributed among left, right, forward, backward. The speed of the users is assigned as a pedestrian between [0,$c_{max}$], where $c_{max}$ denotes the maximum speed of a user. The real-time position of the users is presented by the random walk model in this paper. This model can be replaced by a variety of mobility models such as a deterministic approach and a hybrid approach. In this case, the proposed method to obtain the dynamic movement of the UAVs can be implemented no matter how ground users move. At each timeslot, as the users move, the UAV takes an action ${a_t} \in A$, which also includes choosing a direction for itself, according to the current state, ${s_t} \in S$, based on the decision policy $J$, then, the state of next timeslot is ${s_{t + 1}} = {s_t} + {a_t}$. The q-learning method is utilized to obtain each action at each time slot.

In contrast to the Q-learning based deployment algorithm, the state in the proposed Q-learning based dynamic movement design algorithm to obtain dynamic movement of the UAVs can be divided to two partitions: the dynamic 3D position of the UAVs and the dynamic 2D position of the users, which can be denoted as: $\xi  = (x_{{\text{UAV}}},y_{{\text{UAV}}},h_{{\text{UAV}}},x_{{\text{user}}},y_{{\text{user}}})$. $({x_{{\text{user}}}},{y_{{\text{user}}}})$ is determined by the initial position and movement of the users, $({x_{{\text{UAV}}}},{y_{{\text{UAV}}}},{h_{{\text{UAV}}}})$ is determined by the dynamic of position of the UAVs and the action they take at last time slot.

\begin{remark}\label{remark:convergence}
The convergence rate of the proposed algorithm for solving the problem of the dynamic movement of UAVs varies when the simulation timespan is changed. The smaller the simulation timespan is, the more accurate the dynamic movement will obtain, while the more complex the model will be. There is a tradeoff between improving the QoE of users and the complexity of the proposed algorithm.
\end{remark}

It's worth to be noted that the proposed Q-learning algorithm for the 3D dynamic movement of UAVs is based on the results of the 3D deployment of the UAVs. At the beginning of the time slot, the UAVs are at their optimal position based on the Q-learning based deployment algorithm. Then, at each time slot, the optimal actions of the UAVs are attained based on the mobility of users, which make up the dynamic movement of the UAVs.

\begin{table}[htbp]\footnotesize
\caption{Simulation parameters}
\centering
\begin{tabular}{|c||c||c|}
\hline
\centering
 Parameter & Description & Value\\
\hline
\centering
 $f_c$ & Carrier frequency & 2GHz\\
\hline
\centering
 ${P_{\max }}$ & UAVs' maximum transmit power & 20dBm\\
\hline
\centering
 $N_0$ & Noise power spectral & -170dBm/Hz\\
\hline
\centering
 $U$ & Ground users' number & 100\\
\hline
\centering
 $N$ & UAVs' number & 4\\
\hline
\centering
 $B$ & Bandwidth & 1MHz\\
\hline
\centering
 $b_1$,$b_2$ & Environmental parameters (dense urban) & 0.36,0.21 ~\cite{Mozaffari2017IEEE_J_WCOM}\\
\hline
\centering
 $\alpha $ & Path loss exponent & 2\\
 \hline
\centering
 ${\mu _{LoS}}$ & Additional path loss for LoS & 3dB ~\cite{Mozaffari2017IEEE_J_WCOM}\\
 \hline
\centering
 ${\mu _{NLoS}}$ & Additional path loss for NLoS & 23dB ~\cite{Mozaffari2017IEEE_J_WCOM}\\
 \hline
 \centering
 $\alpha_l, \beta$ & Learning rate, Discount factor & 0.01,0.7\\
 \hline
\end{tabular}
\label{table:parameters}
\end{table}

\section{Numerical Results}
In this section, the numerical results are evaluated to the performance of the QoE of the users with the proposed 3D deployment scheme and dynamic movement obtaining scheme. In this simulation, we consider that 100 users are uniformly and independently distributed within a geographical urban district with size $1{\text{km}} \times 1{\text{km}}$, and the users roam follow random walk mode at each time slot. The other simulation parameters are given in Table~\ref{table:parameters}.

\subsection{Numerical Results of 3D Deployment Problem}

\begin{figure} [t!]
\centering
\includegraphics[width=3in]{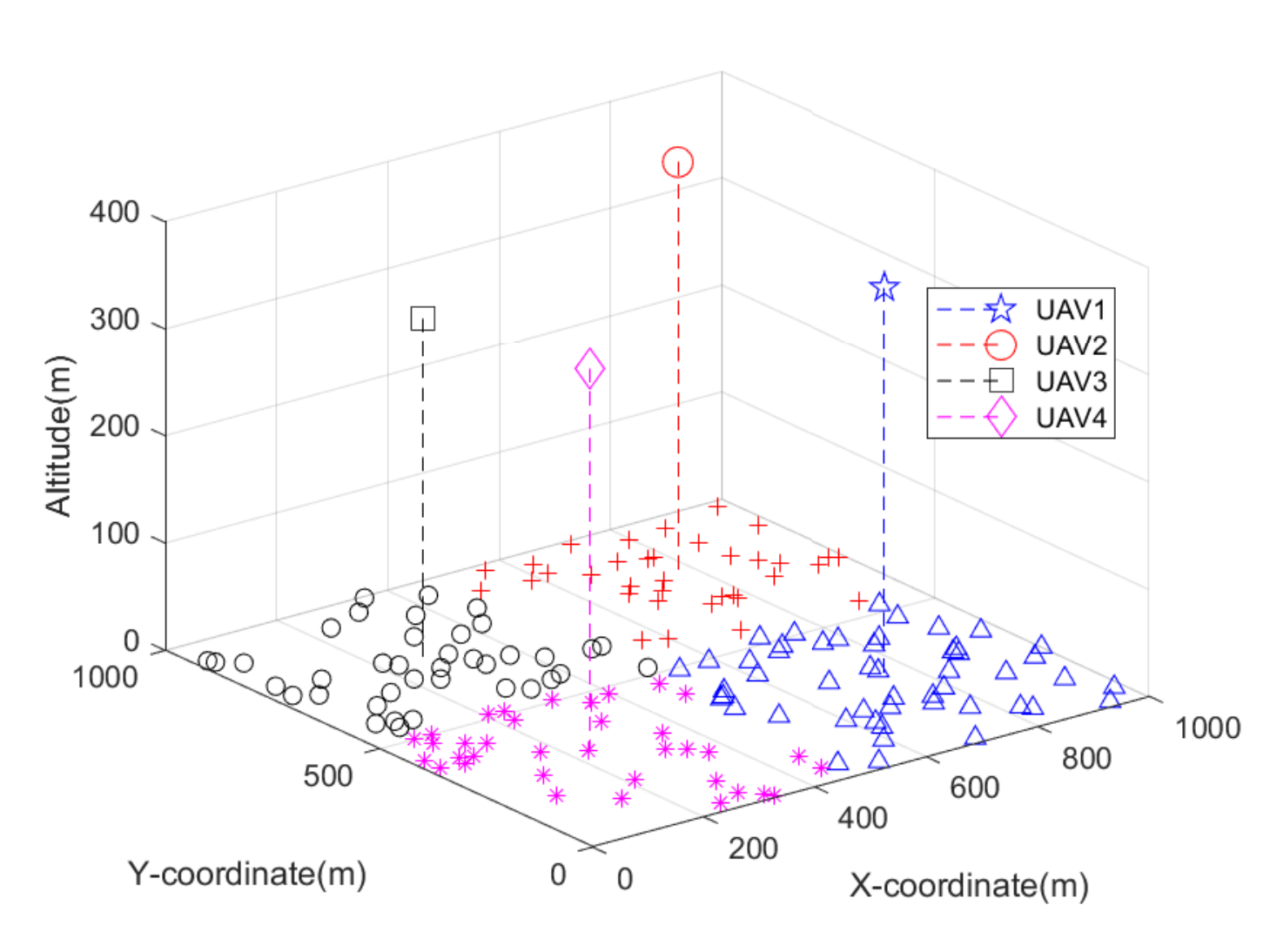}
 \caption{Three dimensional distribution of UAVs and ground users.}\label{3dtu}
\end{figure}

\begin{figure} [t!]
\centering
\includegraphics[width=3in]{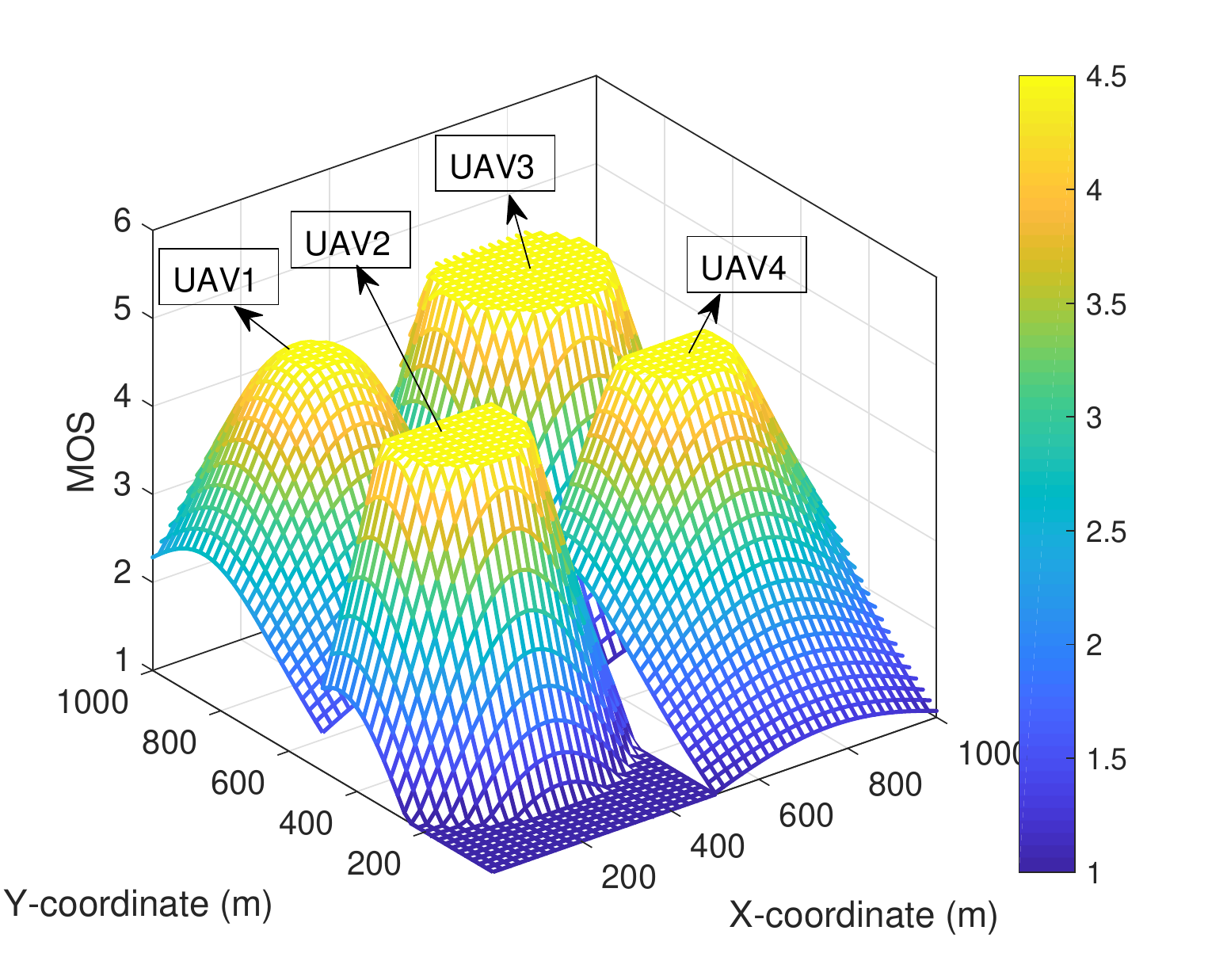}
 \caption{QoE of each ground user.}\label{3dmos}
\end{figure}

Fig.~\ref{3dtu} plots the 3D deployment of the UAVs at the initial time. It can be observed that ground users are divided into 4 clusters asymmetrically, each cluster is served by one UAV. UAVs in different cluster obtains different altitudes. This is because that as the altitudes of the UAVs increase, a higher path loss, and high LoS probability are obtained simultaneously. The altitudes of the UAVs are determined by the density and positions of ground users.

Fig.~\ref{3dmos} plots the QoE of each ground user. It can be observed that as the distance from UAV to a ground user increases, the MOS of this ground user decreases. This result is also analyzed by the insights in \textbf{Remark 1}. In different clusters, the MOS figures are differentiated, this is because ground users have various QoE requirements. In this case, the MOS of a ground user is determined by the distance with UAV as well as the users' QoE requirements.

\begin{figure} [t!]
\centering
\includegraphics[width=3in]{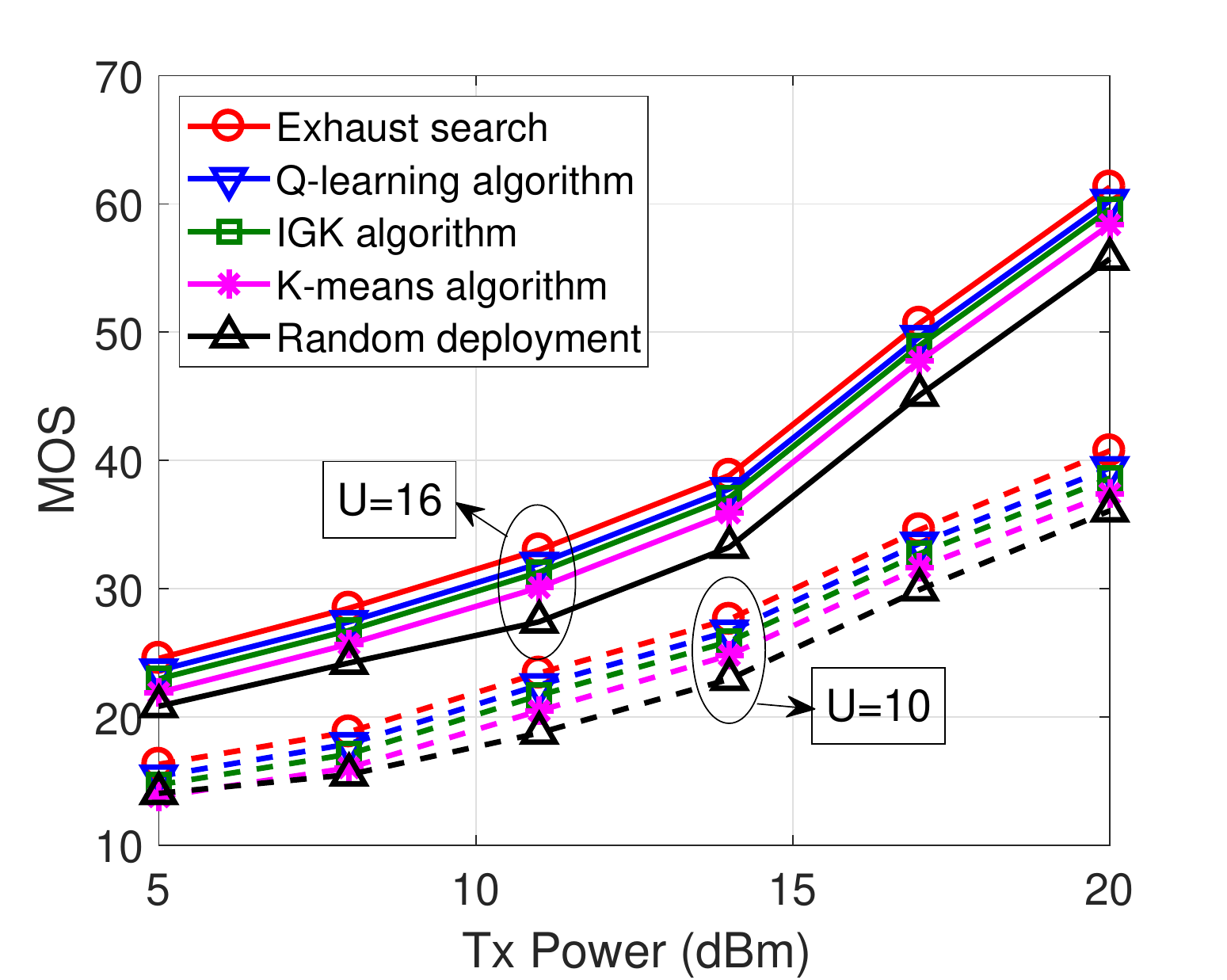}
 \caption{Comparisons of QoE over different algorithms and user number.}\label{mos2}
\end{figure}

\begin{figure} [t!]
\centering
\includegraphics[width=3in]{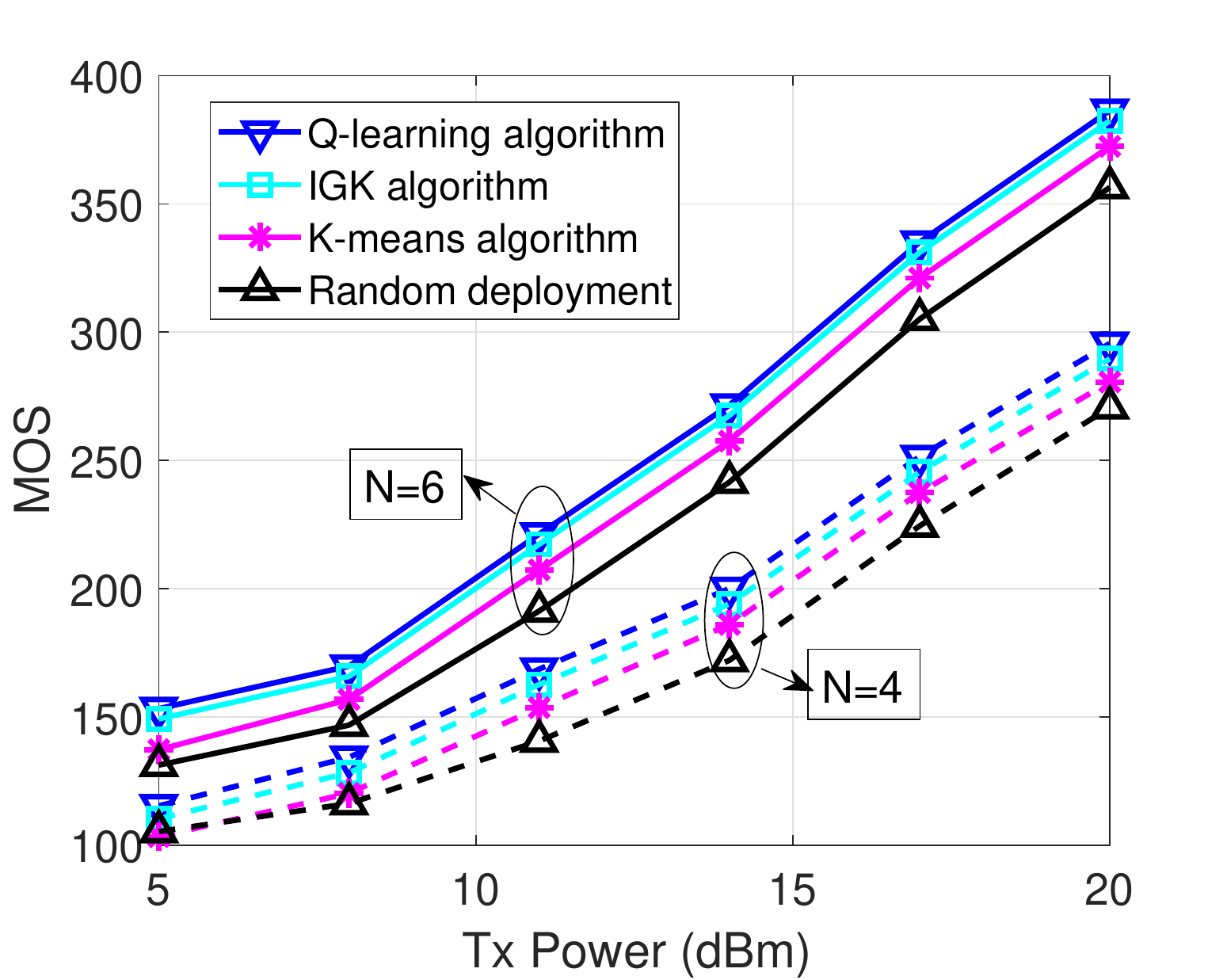}
 \caption{Comparisons of QoE over different algorithms and UAV number.}\label{mos1}
\end{figure}

\begin{figure}
\centering
\includegraphics[width=3in]{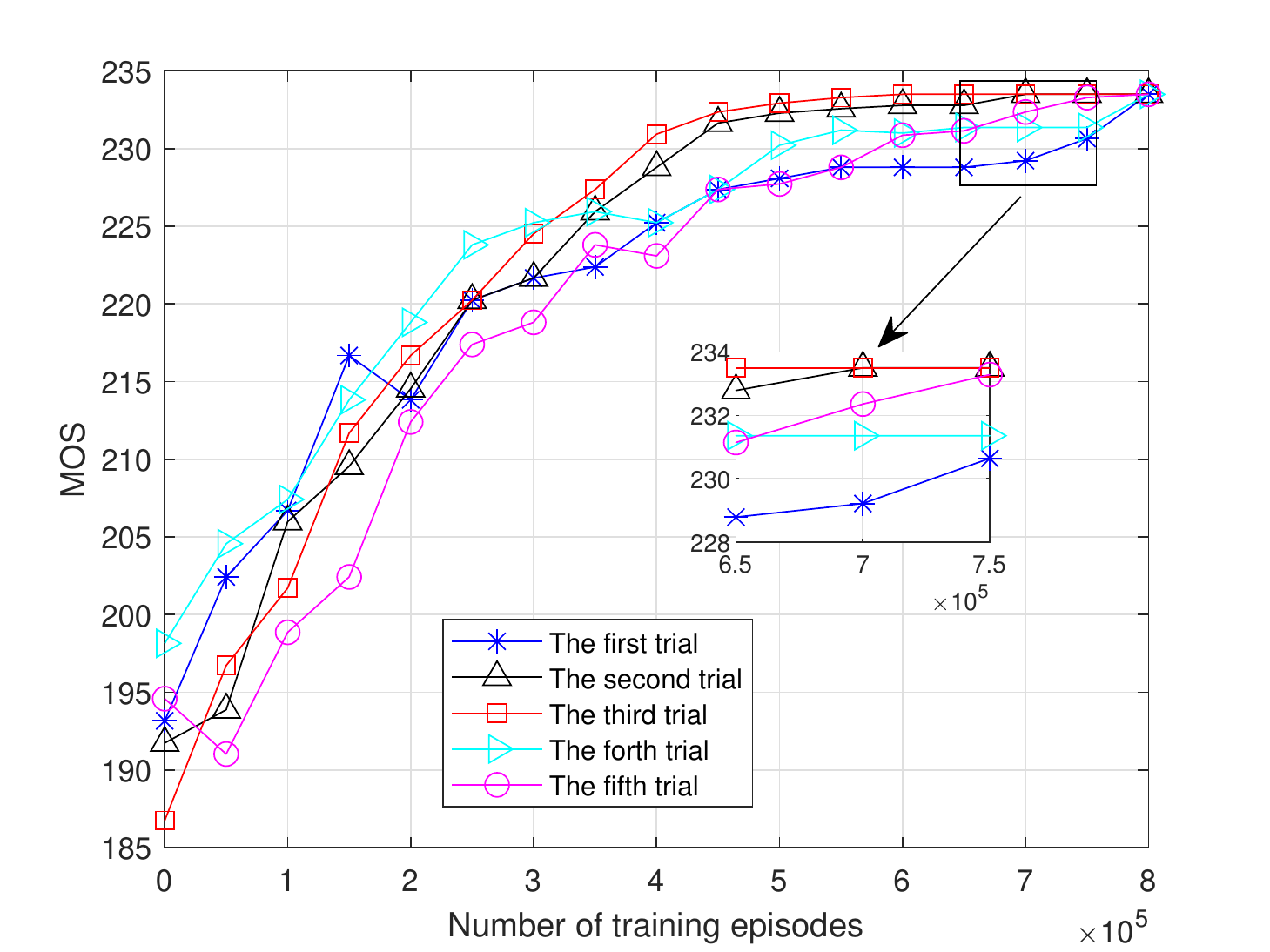}
\caption{Convergence of the proposed algorithm vs the number of training episodes.}\label{convergence}
\end{figure}

Fig.~\ref{mos2} plots the QoE versus the maximum number of users over different algorithms. As the analysis in Section III, the complexity of the exhaust search solution is non-trivial, for the aim of showing the result of an exhaust search solution, a different number of users are considered. From fig.~\ref{mos2}, it can be observed that the total MOS is capable of being improved as the UAVs' transmit power and the ground users' number increase. Additionally, the Q-learning algorithm is capable of achieving better performance than the K-means algorithm and IGK algorithm when increasing the transmit power of UAVs.

Fig.~\ref{mos1} plots the QoE of ground users as the maximum transmit power of UAVs varies. K-means algorithm, IGK algorithm, and random deployment are also demonstrated as benchmarks. One can observe that the sum MOS attained increases when rising to transmit power of UAVs, which is because the received SINR at the users can be improved and improves the sum MOS of ground users. Moreover, it can be observed that increasing the number of clusters is capable of improving the sum MOS. This result is also analyzed by the insights in \textbf{Remark 2}. When the transmit power of UAVs is 5dBm, the MOS of different cluster number is nearly the same. This is because the received SINR is not capable of satisfying the rate requirement of users. It can also be observed from Fig.~\ref{mos1}, increasing the transmit power can further enlarge the gap between the performance of different cluster numbers. Furthermore, the Q-learning algorithm outperforms the K-means algorithm and IGK algorithm while closing to exhaustive search. It is also worth noting that, by training process off-line, when Q-table is updated, the complexity of the Q-learning algorithm and K-means algorithm is the same.

Fig.~\ref{convergence} plots the sum MOS vs the number of training episodes. It can be observed that the UAVs are capable of carrying out their actions in an iterative manner and learn from their mistakes for improving the sum MOS. In this figure, different lines represent different initial positions of UAVs in each trial. It can also be observed that the convergence rate of the proposed algorithm varies when the initial positions of the UAVs are changed. This result is also analyzed by the insights in \textbf{Remark 4}. Despite the initial positions of the UAVs, convergence is achieved after about 450000 episodes.

\subsection{Numerical Results of Dynamic Movement Problem}

\begin{figure} [t!]
\centering
\includegraphics[width=3in]{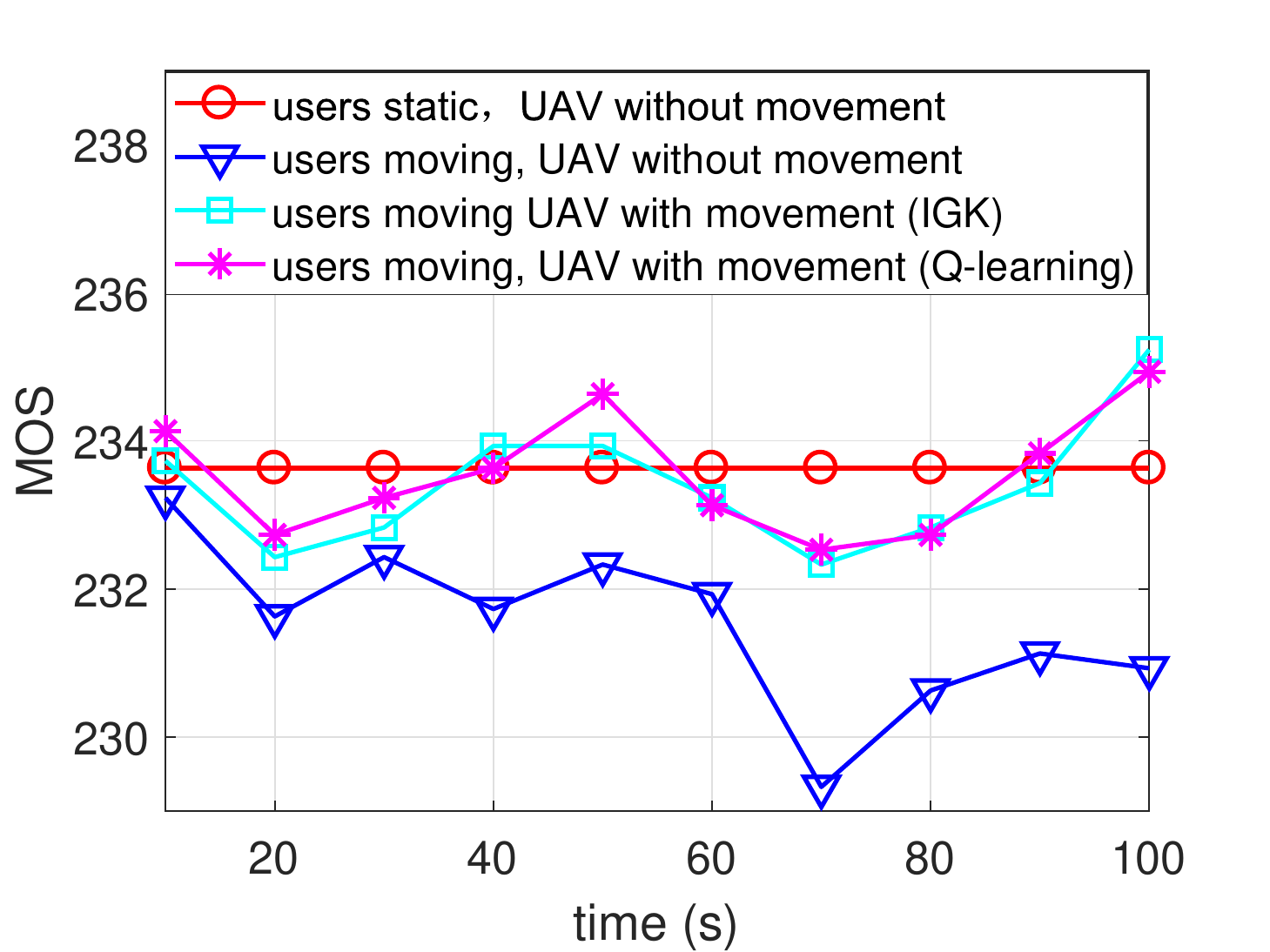}
 \caption{Comparisons of QoE when ground users are moving or static.}\label{finalresult}
\end{figure}

\begin{figure} [t!]
\centering
\includegraphics[width=3in]{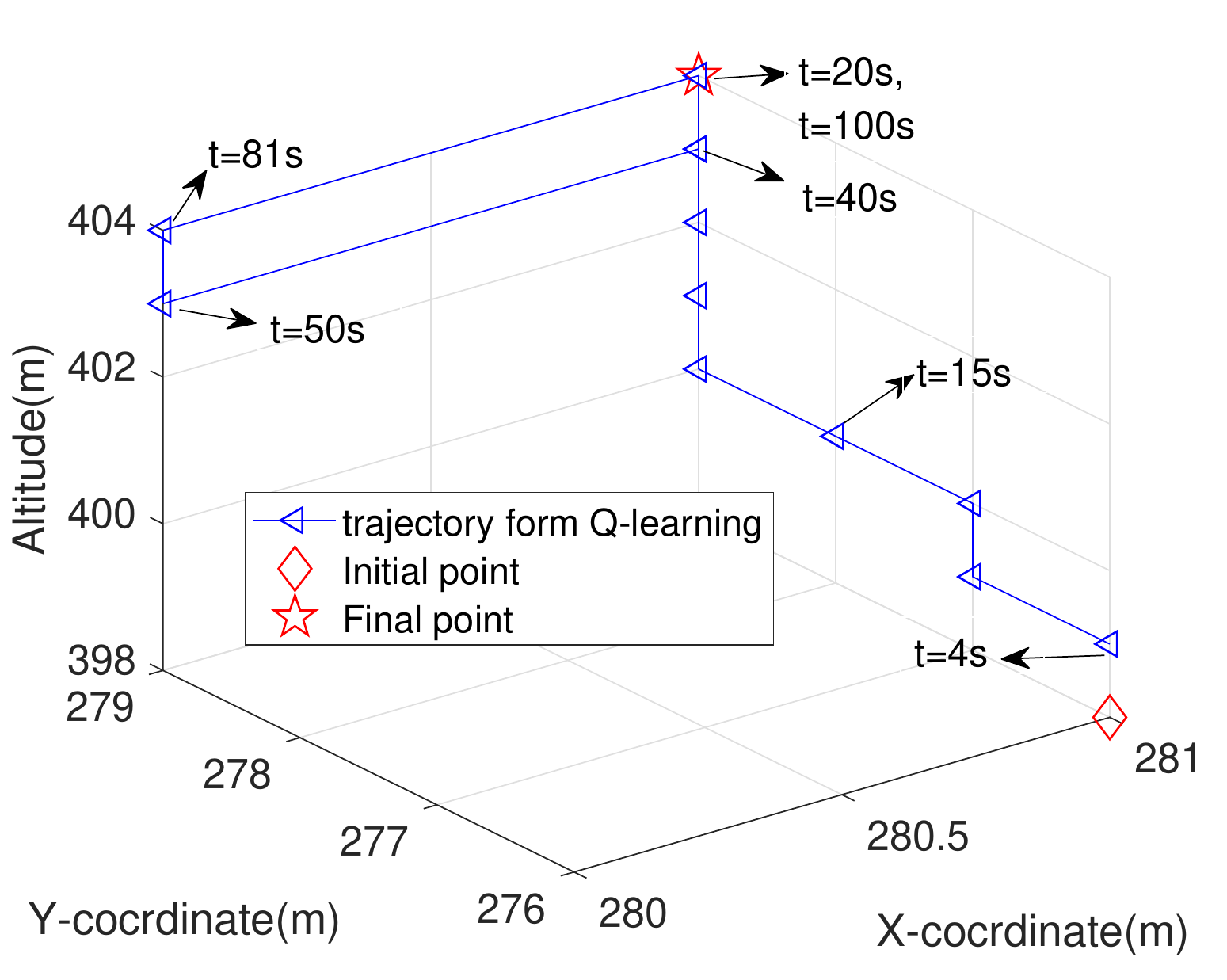}
 \caption{Dynamic movement for one of the four UAVs when ground users are roaming.}\label{trajectory}
\end{figure}

Fig.~\ref{finalresult} plots the sum MOS with the UAVs' movement derived from the Q-learning algorithm. The sum MOS when UAVs are static and the sum MOS with the movement of UAVs derived from the IGK algorithm are also illustrated as a benchmark. When ground users move to follow the random walk model, UAVs are supposed to move along to the movement of users. Otherwise, the sum MOS will decrease as shown in~\ref{finalresult}. This result is also analyzed by the insights in \textbf{Remark 5}. One can also observe that the Q-learning algorithm performs better than the IGK algorithm. Furthermore, note that the computation complexity of the IGK algorithm is non-trivial to obtain dynamic movement of UAVs. In contrast to the IGK algorithm, the Q-learning algorithm enables each UAV gradually to learn its dynamic movement, then the dynamic movement is obtained.

Fig.~\ref{trajectory} plots the dynamic movement of UAVs derived from the proposed approach when ground users move. In this figure, the dynamic movement of a UAV is shown, and the duration time is 100s. In this simulation, we assume all the UAVs can move at a constant speed. At each time slot, UAVs choose a direction from the action space which contains 7 directions, then the dynamic movement will maximize the QoE of ground users at each time slot. It's important to note here, we can adjust the timespan to improve the accuracy of dynamic movement. This, in turn, increases the number of required iterations for convergence. Therefore, a tradeoff exists between improving the QoE of ground users and the running complexity of the proposed algorithm.

\section{Conclusions}

The QoE-driven 3D deployment and dynamic movement of multiple UAVs were jointly studied. The algorithm was proposed for solving the problem of maximizing the sum MOS of the users. By proving the 3D deployment problem as a non-convex problem, three steps were provided to tackle this problem. More particularly, the GAK-means algorithm was invoked to obtain initial cell partition. A Q-learning-based deployment algorithm was proposed for attaining 3D placement of UAVs when users are static. A Q-learning based movement algorithm was proposed for obtaining 3D dynamic movement of UAVs. It is demonstrated that the proposed 3D deployment scheme outperforms the K-means algorithm and IGK algorithm with low complexity. Additionally, with the aid of the proposed approach, the 3D dynamic movement of UAVs is obtained.

\numberwithin{equation}{section}
\section*{Appendix~A: Proof of Lemma 1} \label{Appendix:A}
\renewcommand{\theequation}{A.\arabic{equation}}
\setcounter{equation}{0}

The received SINR of user $k_n$ in time $t$ while connecting to UAV $n$, denoted by ${\Gamma _{{k_n}}}(t)$, can be expressed as
\[{\Gamma _{{k_n}}}(t) = \frac{{{p_{{k_n}}}(t){g_{{k_n}}}(t)}}{{{\sigma ^2}}} \ge \gamma .\tag{A.1}\]
Equation (A.1) can be rewritten as \[
  {P_{\max }} \ge \gamma {\sigma ^2}{K_0}d_{{k_n}}^\alpha (t)({P_{LoS}}{\mu _{LoS}} + {P_{NLoS}}{\mu _{NLoS}}) .\tag{A.2}\]
  It can be easily proved that $({P_{LoS}}{\mu _{LoS}} + {P_{NLoS}}{\mu _{NLoS}}) \le {\mu _{NLoS}}$. When the UAV-user link has a 100\% probability of NLoS connection, the equation will turn to equality.
Thus, we can obtain that \[\begin{array}{l}
{{P_{\max }} \geqslant \gamma {\sigma ^2}{K_0}d_{{k_n}}^\alpha (t){\mu _{NLoS}}}.
\end{array}\tag{A.3} \]
The proof is completed.

\numberwithin{equation}{section}
\section*{Appendix~B: Proof of Proposition 1} \label{Appendix:B}
\renewcommand{\theequation}{B.\arabic{equation}}
\setcounter{equation}{0}

The received SINR of user $k_n$ in time $t$ while connecting to UAV $n$, denoted by ${\Gamma _{{k_n}}}(t)$, can be expressed as Equation (A.1). Equation (A.1) can be rewrite as follows \[{p_{{k_n}}}(t){K_0}^{ - 1}d_{{k_n}}^{ - \alpha }(t){L^{ - 1}} \ge \gamma  {\sigma ^2},\tag{B.1}\]
where $L = {P_{LoS}}({\theta _{{k_n}}}){\mu _{LoS}} + {P_{NLoS}}({\theta _{{k_n}}}){\mu _{NLoS}}$. Then, $d_{{k_n}}^\alpha (t) \le \frac{{{p_{{k_n}}}(t)}}{{\gamma {K_0}{\sigma ^2}L}}\le \frac{{{P_{\max }}}}{{\gamma {K_0}{\sigma ^2}L}}$. In this case, we obtain that  \[\begin{gathered}
  {P_{LoS}}({\theta _{{k_n}}}) = {b_1}{(\frac{{180}}{\pi }{\theta _{{k_n}}} - \zeta )^{{b_2}}}\hfill \\
   {\kern 1pt} {\kern 1pt} {\kern 1pt} {\kern 1pt} {\kern 1pt} {\kern 1pt} {\kern 1pt} {\kern 1pt} {\kern 1pt} {\kern 1pt} {\kern 1pt} {\kern 1pt} {\kern 1pt} {\kern 1pt} {\kern 1pt} {\kern 1pt} {\kern 1pt} {\kern 1pt} {\kern 1pt} {\kern 1pt} {\kern 1pt} {\kern 1pt} {\kern 1pt}  {\kern 1pt} {\kern 1pt}{\kern 1pt} {\kern 1pt} {\kern 1pt} {\kern 1pt} {\kern 1pt} {\kern 1pt} {\kern 1pt} {\kern 1pt} {\kern 1pt} {\kern 1pt} {\kern 1pt} {\kern 1pt} {\kern 1pt} {\kern 1pt} {\kern 1pt} {\kern 1pt} {\kern 1pt} {\kern 1pt} {\kern 1pt} {\kern 1pt} {\kern 1pt} \ge \frac{S}{{({\mu _{LoS}} - {\mu _{NLoS}})}} - \frac{{{\mu _{NLoS}}}}{{{\mu _{LoS}} - {\mu _{NLoS}}}} \hfill \\
\end{gathered},\tag{B.2} \]
where $S = \frac{{{P_{\max }}}}{{\gamma {K_0}{\sigma ^2}d_{{k_n}}^\alpha (t)}}$. Then, we have ${\theta _{{k_n}}}(t) = {\sin ^{ - 1}}(\frac{{{h_n}(t)}}{{{d_{{k_n}}}(t)}}) \ge \frac{\pi }{{180}}[\zeta  + {e^M}]$. Finally, we obtain \[{h_n}(t) \ge {d_{{k_n}}}(t)\sin \left[ {\frac{\pi }{{180}}(\zeta  + {e^M})} \right].\tag{B.4}\]

 Following from (B.4), we have $d_{{k_n}}^{}(t){\kern 1pt}  \le {\left[ {\frac{{{p_{{k_n}}}(t)}}{{\gamma {K_0}{\sigma ^2}L}}} \right]^{{1 \mathord{\left/
 {\vphantom {1 \alpha }} \right.
 \kern-\nulldelimiterspace} \alpha }}}$. It can be proved that ${P_{LoS}}({\theta _{{k_n}}}){\mu _{LoS}} + {\mu _{NLoS}}(1 - {P_{LoS}}({\theta _{{k_n}}}) \le {\mu _{LoS}}$ because of ${\mu _{LoS}} < {\mu _{NLoS}}$. So, we have $d_{{k_n}}^{}(t){\kern 1pt}  \le {(\frac{{{P_{\max }}}}{{\gamma {K_0}{\sigma ^2}{\mu _{LoS}}}})^{{1 \mathord{\left/
 {\vphantom {1 \alpha }} \right.
 \kern-\nulldelimiterspace} \alpha }}}$. It's easy to obtain that ${h_n}(t) \le d_{{k_n}}^{}(t){\kern 1pt} $, then, we have

 \[{h_n}(t){\kern 1pt}  \le {(\frac{{{P_{\max }}}}{{\gamma {K_0}{\sigma ^2}{\mu _{LoS}}}})^{{1 \mathord{\left/
 {\vphantom {1 \alpha }} \right.
 \kern-\nulldelimiterspace} \alpha }}}.\tag{B.5}\]

 The altitude constraint of UAV $n$ is given by
\[{d_{{k_n}}}(t)\sin \left[ {\frac{\pi }{{180}}(\zeta  + {e^M})} \right] \le {h_n}(t){\kern 1pt}  \le {(\frac{{{P_{\max }}}}{{\gamma {K_0}{\sigma ^2}{\mu _{LoS}}}})^{{1 \mathord{\left/
 {\vphantom {1 \alpha }} \right.
 \kern-\nulldelimiterspace} \alpha }}}.\tag{B.6}\]

The proof is completed.

\numberwithin{equation}{section}
\section*{Appendix~C: Proof of Theorem 1} \label{Appendix:C}
\renewcommand{\theequation}{C.\arabic{equation}}
\setcounter{equation}{0}

The sum MOS can be expressed as
\begin{align}\label{mos4}
  \begin{array}{l}
{\rm{ MO}}{{\rm{S}}_{{\rm{total}}}} = \sum\limits_{n = 1}^N {\sum\limits_{{k_n} = 1}^{{{\rm{K}}_n}} {{\rm{MO}}{{\rm{S}}_{k,n}}({r_{{k_n}}})} } {\rm{ }}\\
{\kern 1pt} {\kern 1pt} {\kern 1pt} {\kern 1pt} {\kern 1pt} {\kern 1pt} {\kern 1pt} {\kern 1pt} {\kern 1pt} {\kern 1pt} {\kern 1pt} {\kern 1pt} {\kern 1pt} {\kern 1pt} {\kern 1pt} {\kern 1pt} {\kern 1pt} {\kern 1pt} {\kern 1pt} {\kern 1pt} {\kern 1pt} {\kern 1pt} {\kern 1pt} {\kern 1pt} {\kern 1pt} {\kern 1pt} {\kern 1pt} {\kern 1pt} {\kern 1pt} {\kern 1pt} {\kern 1pt} {\kern 1pt} {\kern 1pt} {\kern 1pt} {\kern 1pt} {\kern 1pt} {\kern 1pt} {\kern 1pt} {\kern 1pt} {\kern 1pt} {\kern 1pt} {\kern 1pt} {\kern 1pt} {\kern 1pt} {\kern 1pt} {\rm{ =  }} - {C_1}\sum\limits_{n = 1}^N {\sum\limits_{{k_n} = 1}^{{{\rm{K}}_n}} {\ln (d({r_{{k_n}}}))} }  + {C_2}.
\end{array}
\end{align}

Equation (C.1) can be represented and given by \eqref{Pout_2} at the top of next page.
\begin{figure*}
\begin{align}\label{Pout_2}
{\text{ MO}}{{\text{S}}_{{\text{total}}}} = {C_2} - {C_1}\sum\limits_{n = 1}^N {\sum\limits_{{k_n} = 1}^{{{\text{K}}_n}} {\ln ({\text{3}}RTT + \frac{{FS}}{{{r_{{k_n}}}}} + L(\frac{{MSS}}{{{r_{{k_n}}}}} + RTT){\kern 1pt}  - \frac{{2MSS({2^L} - 1)}}{{{r_{{k_n}}}}})}}.
\end{align}
\end{figure*}

The instantaneous achievable rate of user ${{r_{{k_n}}}}$ in (B.1) can be expressed as \[\begin{array}{l}
{r_{{k_n}}}(t) = {B_{{k_n}}}{\log _2}(1 + \frac{{{p_{{k_n}}}(t){g_{{k_n}}}(t)}}{{{\sigma ^2}}})\\
 = \frac{B}{{{K_n}}}{\log _2}(1 + \frac{{\frac{{{P_{\max }}}}{{{K_n}{{(\frac{{4\pi {f_c}}}{c})}^2}d_{{k_n}}^\alpha (t)[{P_{{\rm{LoS}}}}{\mu _{{\rm{LoS}}}} + {P_{{\rm{NLoS}}}}{\mu _{{\rm{NLoS}}}}]}}}}{{{\sigma ^2}}}).
\end{array}\tag{C.3}\]

It is easy to verify that $f(z) = \log (1 + \frac{\gamma }{A })$ is convex with regard to $A > 0$, $\gamma > 0$ and $\gamma  > A$. Then, the non-convexity of (C.3) equals to the non-convexity of $d_{{k_n}}^\alpha (t)({P_{{\text{LoS}}}}{\mu _{{\text{LoS}}}} + {P_{{\text{NLoS}}}}{\mu _{{\text{NLoS}}}})$.

It can be proved that $d_{{k_n}}^\alpha (t)({P_{{\text{LoS}}}}{\mu _{{\text{LoS}}}} + {P_{{\text{NLoS}}}}{\mu _{{\text{NLoS}}}})$ is convex over the altitude of UAVs. However, as we have three variate (X-coordinate, Y-coordinate and altitude of UAVs), and ${x_n}(t)$'s, ${y_n}(t)$'s and ${h_n}(t)$'s vary at each time slot, $d_{{k_n}}^\alpha (t)({P_{{\text{LoS}}}}{\mu _{{\text{LoS}}}} + {P_{{\text{NLoS}}}}{\mu _{{\text{NLoS}}}})$ is non-convex.In this case, (C.3) and (C.1) is also non-convex over ${x_n}(t)$'s, ${y_n}(t)$'s and ${h_n}(t)$'s.

The proof is completed.

\numberwithin{equation}{section}
\section*{Appendix~D: Proof of Theorem 2} \label{Appendix:D}
\renewcommand{\theequation}{D.\arabic{equation}}
\setcounter{equation}{0}

In order to prove that the problem (13a) is NP-hard, three steps are supposed to be taken based on the computational complexity theory. Firstly, a known NP-complete decision problem $Q$ is supposed to be chosen. Secondly, a polynomial time transformation from any instance of $Q$ to an instance of the problem (13a) is supposed to be constructed. Thirdly, the two instances are supposed to be proven that they have the same objective value.

In this article, we prove that (13a) is NP-hard even the QoE requirements of ground users are the same, in the meantime, the probability of LoS connections from UAVs towards ground users is one. Thus, problem (13a) can be rewritten as \[\begin{gathered}
  \mathop {\max }\limits_{C{\text{,}}Q,H} {\kern 1pt} {\kern 1pt} {\kern 1pt} {\kern 1pt} {\kern 1pt} {\kern 1pt} {\kern 1pt} {\kern 1pt} {\kern 1pt} {\kern 1pt} {\kern 1pt} {\kern 1pt} {\kern 1pt} {\kern 1pt} {\kern 1pt} {\kern 1pt} {\kern 1pt} {\kern 1pt} {\kern 1pt} {R_{sum}} \hfill \\
  {\text{s}}{\text{.t}}{\text{.}}{\kern 1pt} {\kern 1pt} {\kern 1pt} {\kern 1pt}{\kern 1pt} {\kern 1pt} {\kern 1pt} {\kern 1pt} {\kern 1pt} {\kern 1pt} {\kern 1pt} {\kern 1pt} {\kern 1pt} {\kern 1pt} {\kern 1pt} {\kern 1pt} {\kern 1pt} {\kern 1pt} {\kern 1pt} {\kern 1pt} {\kern 1pt} {\kern 1pt} {\kern 1pt} {13(b),13(c),13(d),13(e),13(f)}. \hfill \tag{D.1}\\
\end{gathered} \]

As the probability of LoS connections is one, the overall achievable sum rate is \[\begin{array}{l}
{R_{{\rm{sum}}}} = \sum\limits_{n = 1}^N {\sum\limits_{{k_n} = 1}^{{{\rm{K}}_n}} {{r_{{k_n}}}(t)} } \\
 = \sum\limits_{n = 1}^N {\sum\limits_{{k_n} = 1}^{{{\rm{K}}_n}} {\frac{B}{{{K_n}}}{{\log }_2}\left( {1 + \frac{{{P_{\max }}}}{{{\sigma ^2}{K_n}{{(\frac{{4\pi {f_c}}}{c})}^2}d_{{k_n}}^\alpha (t){\mu _{{\rm{LoS}}}}}}} \right)} }
\end{array}.  \tag{D.2} \]

Following from (D.2), the overall achievable sum rate only relate to the sum Euclidean distance. Furthermore, when the UAVs' altitudes are assumed to be fixed, the problem (D.1) is simplified as \[\begin{gathered}
  \mathop {\min }\limits_{C{\text{,}}Q} {\kern 1pt} {\kern 1pt} {\kern 1pt} {\kern 1pt} {\kern 1pt} {\kern 1pt} {\kern 1pt} {\kern 1pt} {\kern 1pt} {\kern 1pt} {\kern 1pt} {\kern 1pt} {\kern 1pt} {\kern 1pt} {\kern 1pt} {\kern 1pt} {\kern 1pt} {\kern 1pt} {\kern 1pt} \sum\limits_{n = 1}^N {\sum\limits_{{k_n} = 1}^{{{\rm K}_n}} {{d_{{k_n}}}} }  \hfill \\
  {\text{s}}{\text{.t}}{\text{.}}{\kern 1pt} {\kern 1pt} {\kern 1pt} {\kern 1pt} {\kern 1pt} {\kern 1pt} {\kern 1pt} {\kern 1pt} {\kern 1pt} {\kern 1pt} {\kern 1pt} {\kern 1pt} {\kern 1pt} {\kern 1pt} {\kern 1pt} {\kern 1pt} {\kern 1pt} {\kern 1pt} {\kern 1pt} {13(b),13(d)} .\hfill \\\tag{D.3}\\
\end{gathered} \]

Problem (D.3) is known as the planar K-means problem. The authors in~\cite{mahajan2012planar} demonstrate the NP-hardness of planar K-means problem by a reduction from the planar 3-SAT problem, which is known as an NP-complete problem. The proving process will not be shown in this paper, and it can be seen in~\cite{mahajan2012planar}. Note that the problem (D.3) is a special case of problem (13a), then the original problem in (13a) is NP-hard.

The proof is completed.

\bibliographystyle{IEEEtran}
\bibliography{mybib}

\end{document}